# Transitions between superstatistical regimes: validity, breakdown and applications

Petr Jizba[a,b], Jan Korbel[c,a], Hynek Lavička[d,e], Martin Prokš[a], Václav Svoboda[a], Christian Beck[f]

[a] Faculty of Nuclear Sciences and Physical Engineering, Czech Technical University in Prague, Břehová 7, 11519, Prague, Czech Republic
[b] Institute of Theoretical Physics, Freie Universität in Berlin, Arnimallee 14, 14195 Berlin, Germany
[c] Department of Physics, Zhejiang University, Hangzhou 310027, P. R. China
[d] Alten Belgium N.V., Chausse de Charleroi 112, 1060 Brussels, Belgium
[e] Department of Institutional, Environmental and Experimental Economics, University of Economics in Prague, Square W. Churchilla 4, 130 67 Praha 3, Czech Republic
[f] Queen Mary University of London, School of Mathematical Sciences, Mile End Road, London E1 4NS, United Kingdom

**Abstract**

Superstatistics is a widely employed tool of non-equilibrium statistical physics which plays an important rôle in analysis of hierarchical complex dynamical systems. Yet, its "canonical" formulation in terms of a single nuisance parameter is often too restrictive when applied to complex empirical data. Here we show that a multi-scale generalization of the superstatistics paradigm is more versatile, allowing to address such pertinent issues as transmutation of statistics or inter-scale stochastic behavior. To put some flesh on the bare bones, we provide a numerical evidence for a transition between two superstatistics regimes, by analyzing high-frequency (minute-tick) data for share-price returns of seven selected companies. Salient issues, such as breakdown of superstatistics in fractional diffusion processes or connection with Brownian subordination are also briefly discussed.

## 1. Introduction

The concept of "emergence" plays an important role in modern statistical physics. One of the key characteristics of emergence is that the observed macroscopic-scale dynamics and related degrees of freedom differ drastically from the actual underlying microscopic-scale physics [1, 2, 3]. Such systems are often characterized by hierarchical structures of underlying dynamics. Superstatistics provides an explicit realization of this paradigm: It posits that the emergent behavior can be in many cases regarded as a superposition of several statistical systems that operate at different spatio-temporal scales [4]. The essential assumption of the superstatistics scenario is the existence of different spatio-temporal scales which are largely separated from each other, so that the system has enough time to relax to a local equilibrium state and to stay within it for some (phenomenologically relevant) time. The most common superstatistics applications are concerned with two characteristic scales only. In this framework a broad range of successful applications has been recently reported; these include hydrodynamic turbulence [5], non-stationary dynamical processes with time-varying multiplicative noise exponents [6], turbulence in quantum liquids [7], models of

*Email addresses:* p.jizba@fjfi.cvut.cz (Petr Jizba), korbeja2@fjfi.cvut.cz (Jan Korbel), lavicka@fjfi.cvut.cz (Hynek Lavička), proks@fjfi.cvut.cz (Martin Prokš), svoboda@fjfi.cvut.cz (Václav Svoboda), c.beck@qmul.ac.uk (Christian Beck)

the metastatic cascade in cancerous systems [8], complex networks [9], ecosystems driven by hydro-climatic fluctuations [10], pattern-forming systems [11], wind velocity fluctuations [12, 13], share price fluctuations [14, 15] and high-energy physics [16, 17].

In their recent paper [18], D. Xu and C. Beck have brought yet another twist into the superstatistics paradigm by suggesting that in certain financial time series one can observe a temporal breakdown from the log-normal-superstatistics (valid on the minute timescales) to the Gamma-superstatiscs (valid on the daily timescales). This scale-dependent "transmutation" of statistics is a very interesting observation which is in many respects akin to a similar behavior known from the theory of continuous phase transitions. There the transmutation of statistics is imprinted in the behavior of the ensuing two-point auto-correlation function which in the disordered phase (above the critical temperature $T_c$) decays exponentially, while at the critical point it "transmutes" to a power-law decay. The latter signalizes the long-range correlated behavior that may lead to an infinite second (or even first) moment. In fact, the divergence of the lowest moments (as know, for instance, for certain Lévy distributions) implies the absence of underlying physical scales. In statistical physics, the absence of physical scales is interpreted as *scale invariance* which in turn invokes the notion of self-similarity which is a typical hallmark of the presence of a (stable) fixed point in the state space. The primary aim of this paper is to promote and elaborate the issue of "transmutation" of statistics in the superstatistics framework from the viewpoint of infinitely-divisible distributions and theory of critical phenomena. Our considerations will be bolstered with some explicit illustrations from financial market.

The structure of the paper is as follows. To set the stage we elucidate in the next section the inner workings of the superstatistics paradigm. We also outline potential generalizations of the "canonical" superstatistics scenario of Beck *et al.* [4, 20] by considering more characteristic scales and different stable distributions for *prior*. In Section 3, we provide a numerical evidence for a transition between two superstatistics regimes, by analyzing high-frequency (minute-tick) data for share-price returns of seven selected companies. In doing so, we first employ the Multifractal Detrended Fluctuation Analysis (MFDFA) and Surrogate MFDFA to identify within each of the seven time series two (well separated) time scales with qualitatively different underlying dynamics. In the second step, we use the maximum likelihood method together with Kolmogorov–Smirnov, Cramér–von Mises and Anderson–Darling distribution distances to fit optimally the two scale statistical behaviors with existent universality classes in superstatistics. We show that four (out of seven chosen) share-price return time series can be quantitatively well understood as a transition between two superstatistics regimes. In particular, in all four cases we observed a transition from the log-normal-superstatistics (on ~50 minute scale) to Gamma-superstatistics (on ~400 minute scale). Some mechanisms of the formation of regime switching between different superstatistics are briefly discussed in Section 4. There we first comment on Xu–Beck's *synthetic model* and then show that an alternative explanation can be provided via multi-scale superstatistics with the help of the renormalization-group technique. In Section 5, we discuss some potential pitfalls that might happen in the data analysis when the superstatistics paradigm is used too naively or uncritically. Finally, Section 6 summarizes our results and discusses possible extensions of the present work. For the reader's convenience, we give in Appendix A a brief glossary of the companies whose share-price returns are considered in the text. The paper is also accompanied by the Supplementary Material which collects additional supporting material not present but referred to in the main text.

## 2. Brief review of "canonical" superstatistics

In this section we briefly review some of the essentials from superstatistics that will be needed in the main body of the text. Following Refs. [4, 20, 21], we consider an intensive parameter $\beta \in [0, \infty)$ that appreciably changes over time scales that are much larger than the typical relaxation time of the local dynamics. The random variable $\beta$ can be in practice identified, e.g., with the inverse temperature [4, 20, 21], energy dissipation rate (turbulent flow in Kolmogorov theory) [22], volatility (econophysics) [15], einbein (quantum relativistic particles) [16, 17], etc. On an intuitive ground, one may understand the superstatistics by using the *adiabatic* Ansatz. Namely, the system under consideration, during its evolution, travels within its state space $\mathcal{M}$ (described by a state variable $A \in \mathcal{M}$) which is partitioned into small cells characterized by a sharp value of some intensive parameter $\beta$. Within each cell, the system is described by the conditional distribution $\wp(A|\beta)$. As $\beta$ varies adiabatically from cell to cell according to some *mixing* (or *smearing*) distribution $f(\beta)$, the joint distribution of finding the system with a sharp value of $\beta$ in the state $A$ is $\wp(A, \beta) = \wp(A|\beta) f(\beta)$, which is nothing but the De Finetti–Kolmogorov relation. The resulting macro-scale (emergent) statistics $\wp(A)$ for finding system in the state $A$ is obtain by eliminating the nuisance parameter $\beta$ through marginalization, that is

$$\wp(A) \ = \ \int_0^\infty \mathrm{d}\beta \ \wp(A|\beta) f(\beta) \, . \tag{1}$$

The macro-scale distribution $\wp(A)$ is known as the *marginal distribution*, while the local-equilibrium distribution $\wp(A|\beta)$ is the so-called prior distribution, or simply *prior*. In addition, superstatistic as defined in Eq. (1) mathematically qualifies as a form of *slow modulation* [26].

By assuming that the local equilibrium is caused by small independent random fluctuations (such as random collisions of molecules in gases or random buy/sell decisions of individual stock investors) one can identify prior $\wp(A|\beta)$ with some *stable distribution*. In the "canonical" superstatiscs one assumes that the local equilibrium is a consequence of the standard (i.e., Lindeberg or Lyapunov type) central-limit theorem [23]. For this reason the prior is typically equal to the Gaussian distribution. Characteristic example is provided by the Maxwell–Boltzmann velocity distribution in D dimensions

$$\wp(\mathbf{u}|\beta) \ = \ \left(\frac{m\beta}{2\pi}\right)^{D/2} \exp\left(-\frac{1}{2}\beta m \mathbf{u}^2\right) , \tag{2}$$

where $A \equiv \mathbf{u}$ and $m$ are a particle speed and mass, respectively. The intensive parameter $\beta = 1/k_B T$ is in this case related to the inverse absolute temperature $T$.

While in principle any smearing probability density function (PDF) $f(\beta)$ is possible in the superstatistics approach, in practice one usually observes only a few relevant PDF's. These are the *Gamma*, *inverse Gamma* and *log-normal* distribution. In other words, in complex systems with (two-)time-scale separation one usually observes three physically relevant "universality classes" [30]: a) Gamma-superstatistics (leading to Tsallis thermostatistics [54]), b) inverse Gamma-superstatistics and c) log-normal-superstatistics. The reason for these classes can be traced, according to Beck *et al.* [30], to three typical phenomenological situations that could be realistically responsible for emergence of the random variable $\beta$.

The essential assumption of the superstatistics scenario is the existence of sufficient spatio-temporal scale separations between relevant dynamics within the studied system. In the "canonical" superstatistics, cf. Eq. (1), one typically confines itself to two characteristic scales only. On the other hand, one can employ (at least in principle) more general marginalization procedure by introducing more nuisance parameters $\beta_i$. In particular, we wish to consider a non-equilibrium system with a well defined local equilibrium state $A$ and a set of slowly fluctuating intensive parameters $\{\beta\}_{i=1}^n$. While the local equilibrium scale $s_0$ describes a typical relaxation time of the local dynamics, a hierarchy of well separated time scales $\{s_i\}_{i=1}^n$ characterizes fluctuation scales of $\{\beta_i\}_{i=1}^n$. In spirit of superstatistics we assume that the scales are ordered so that $s_0 \ll s_i \ll s_{i+1}$. With this we can write the joint distribution of finding the system with sharp values of $\{\beta_i\}_{i=1}^n$ in the state $A$ as

$$\begin{aligned} \wp(A,\beta_1,\beta_2,\ldots,\beta_n) \ &= \ \wp(A|\beta_1,\beta_2,\ldots,\beta_n)\wp_0(\beta_1,\beta_2,\ldots,\beta_n) \\ &= \ \wp(A|\beta_1,\beta_2,\ldots,\beta_n)\wp_1(\beta_1|\beta_2,\ldots,\beta_n)\wp_2(\beta_2|\beta_3,\ldots,\beta_n)\cdots\wp_{n-1}(\beta_{n-1}|\beta_n)f(\beta_n) \, , \end{aligned} \tag{3}$$

where on the second line the De Finetti–Kolmogorov relation was employed recursively. Clearly, it is in practice hopeless to control PDFs at all scales required (fine tuning problem). At best, one can control only few adjacent scales and ensuing relative changes in dynamics they describe. This allows to pass to a Markovian approximation in each conditional PDF in (3). Then, because of the Markov property the marginal PDF takes the form

$$\wp(A) \ = \ \int \mathrm{d}\beta_1 \ldots \mathrm{d}\beta_n \ f(\beta_n) \prod_{i=1}^{n-1} \wp_i(\beta_i|\beta_{i+1})\wp(A|\beta_1,\beta_2,\ldots,\beta_n) \, . \tag{4}$$

Note, that $\wp(\beta_i|\beta_{i+1})$ effectively describes the smearing PDF on the $s_i$-th scale. By writing $\wp(A|\beta_1,\beta_2,\ldots,\beta_n)$ in (4) we wish to emphasize that $A$ might, in general, depend on all considered scales.[1] It is quite typical, that when

[1]For instance, when $A$ would represent stock-market returns $r$ then the shortest characteristic time scale $s_0$ would be few trading minutes (e.g., for S&P 500 the autocorrelation function of $r$ is around 4 minutes). In this respect $r$'s are basically uncorrelated random variables, nevertheless, they are not independent since higher-order correlations reveal a richer structure — the autocorrelation functions of non-linear functions of $r$ define entire hierarchy of characteristic time scales (e.g., for S&P 500 the characteristic decorrelation time of $|r|$ or $r^2$ spans several months and for $r^4$ it is around two years).

the time scale $s_0$ for prior is very short, then the large-amplitude, temporally correlated fluctuations will start to be relevant, and so in order to describe the presumed local equilibrium, which is a core assumption in superstatistics, one should consider various generalized central limit theorems (CLTs); be it the CLT of Lévy and Gnedenko [23, 24] for non-Gaussian stable distributions (where infinite variances are allowed) or diverse CLTs for correlated random variables [25]. Thus, for the short-scale prior $\wp(A|\beta_1)$ it might be more appropriate to consider a stable PDF different from the Gaussian one.

## 3. Empirical evidence for transmutation of statistics within superstatistics paradigm

In this section we will provide a direct empirical evidence for the existence of regime switching between two different superstatistical regimes. In particular, our analysis will focus on selected financial time series where a non-trivial time-scale-dependent market dynamics might be viably conceived. In fact, markets consist of a number of agents working in different time horizons. It is therefore natural to expect that the dynamics of the interrelations between markets consist of scales that possibly behave differently. Indications of this kind of structural behavior can be conveniently studied with the help of superstatistics. Here we will see two typical scenarios: a) different-scale market dynamics behind times series can be described by two equal-type superstatistics with different parameters, b) different-scale market dynamics are described with two different superstatistics (transmutation of statistics). To this end we will employ the data sets of share-price returns of seven companies from different sectors recorded on the minute-tick basis during period from the 2nd January, 1998 to 22nd May, 2013. For the list of companies used and associated glossary see Appendix A.

### *3.1. Signatures of statistics transmutation*

In order to find time scales characterizing different underlying dynamics for the data sets at hand we will employ here MFDFA and Surrogate MFDFA analyses. This will serve two purposes; a) it will allow to identify clean quantitative signatures of distinctive scale-dependent dynamics which could potentially represent transition between superstatistics regimes, b) it will serve as an independent check of results obtained via superstatistics means in the following subsection.

When analyzing financial time series of stock prices $\{S_i\}$ the typical quantities of interest are log-returns

$$r_i = \log \frac{S_{i+1}}{S_i} . \tag{5}$$

In our case $\{r_i\}$ represent increments of logarithm of stock price during sampling time $\tau = 1$ minute, which corresponds to time scale at which the stock prices are recorded.

Originally, Detrended Fluctuation Analysis (DFA) was introduced in Refs. [42, 43, 44, 45, 46]. Following developments improved the method and Multifractal Detrended Fluctuation Analysis was proposed in Ref. [47]. Relation of MFDFA with alternative methods of time-series analysis were discussed in a number of papers, see e.g., Ref. [48] and citation therein.

MFDFA is, as a rule, time consuming analysis which needs to be implemented efficiently in order one could investigate large datasets with a sufficient precision. The approach implemented in this work was developed in Refs. [39, 40]. In particular, the dataset is divided into samples $X^{seg,w}$ of fixed size $s$ where $w \in \{1,\ldots,N_s\}$ and within each sample a local trend $X^{prof,w}$ is formed by a polynomial of fixed order $o$. We calculate sample variance of deviations from a trend

$$F^2(s,w) = \frac{1}{s}\sum_{i=1}^{s}\left(X^{seg,w}(i) - X^{prof,w}(i)\right)^2 . \tag{6}$$

The Fluctuation function of the $q$-th order is defined as

$$F_q(s) = \begin{cases} \text{for } q \neq 0, & \left\{\frac{1}{N_s}\sum_{w=1}^{N_s}\left[F^2(s,w)\right]^{q/2}\right\}^{1/q} , \\ \text{for } q = 0, & \exp\left[\frac{1}{2N_s}\sum_{w=1}^{N_s}\ln\left(F^2(s,w)\right)\right] . \end{cases} \tag{7}$$

If the series is long-range power correlated, then the fluctuation function will show power-law behavior $F_q(s) \sim s^{h(q)+n}$. Here $h(q)$ stands for the *generalized* Hurst exponent and $n \in \mathbb{Z}$ represents initial pre-processing using integration (positive $n$) or derivation (negative $n$) of the dataset. The reason for the preprocessing is the location of $h(q)$ in a region close to 0, for Gaussian noise with long-range correlations it must hold $-\frac{1}{2} \leq h(q) \leq \frac{1}{2}$. On the other hand, real data often need to have shifted $h(q)$. In our case we did not pre-process the dataset and thus $n = 0$[2]. We note that the power law may not hold in the whole range of $s$ but it may hold piecewise giving rise to the generalized Hurst exponents in a region. Results of the analysis does not qualitatively depend on the order $o$ of the detrending polynomial, see Fig. 2 and 4. Minor quantitative deviations are present in the analysis but they do not interfere with robustness of conclusions presented.

If the aforementioned scaling of $F_q(s)$ exists then $\ln F_q(s)$ will depend linearly on $\ln s$, with $h(q)$ representing the slope. A *monofractal* time series is characterized by unique $h(q)$ for all values of $q$. In general $h(q)$ depends non-trivially on $q$. For stationary time series $h(2)$ equals to the *conventional* Hurst exponent $H$.

The exponent $h(q)$ is related to the classical correlation exponent $\tau(q)$ by the relation $\tau(q) = q \cdot h(q) - 1$. Monofractal series with long-range correlations are characterized by $\tau(q)$ that are linearly dependent on $q$ with a single Hurst exponent $H$. Multifractal signal have multiple Hurst exponent $h(q)$ and $\tau(q)$ depends on $q$ non-linearly.

The Hurst exponent $h(q)$ is related with the singularity spectrum $f(\pi)$ and the correlation exponent $\tau(q)$ by

$$f(\pi) = q(\pi) \cdot \pi - \tau(q(\pi)) , \tag{8}$$

$$\pi = \frac{d\tau(q)}{dq} = q \cdot h'(q) + h(q) . \tag{9}$$

Here $\pi$ is the Hölder–Lipschitz exponent (or the singularity strength) and $f(\pi)$ specifies the Hausdorff dimension of the subset series that is characterized by a fixed value of $\pi$. The multifractal spectrum provides information about relative importance of various fractal exponents in the series, e.g., the width of the spectrum denotes range of exponents.

To investigate intrinsic properties of the time series we use shuffled and surrogate datasets obtained by random shuffling of the dataset and randomization of phases in the ensuing Fourier spectrum. We define the shuffled Hurst exponent $h_q^{shuf}$ and surrogate Hurst exponent $h_q^{sur}$ that are used to obtain properties of correlations and non-linear properties (for more details see Ref. [39]). We might notice that our analysis reveals an existence of two characteristic time scales; a) few tens of minutes and b) around hundred of minutes. There seems to be also higher time scales — over 2500 minutes, but we do not consider them in the present analysis.

### 3.2. Analysis based on supertatistics

As elaborated in previous section and the Supplementary Material [50] it is reliable to consider a sampling interval to be at least 20 minutes. In addition, as seen in [50] this sampling interval is adequate for all seven time series. Consequently, we consider new time series constructed as[3]

$$r_j^{(20)} = \sum_{i=1}^{20} r_{i+(j-1)20}^{(\text{min})} , \qquad j \in \{1, \ldots, \lfloor N/20 \rfloor\} , \tag{10}$$

where $\lfloor \cdots \rfloor$ is the floor function and $N$ is the length of the original series sampled every minute. For simplicity's sake we suppress in following the superscript indicating scale of the time series and use only 20-min scale data unless stated otherwise. In paper [18] such adjustments were not done and that might be the reason for slightly different conclusions.

With the reliable data at hand the next step is to determine time interval in which intensive parameter stays constant. Here we follow the approach presented in Ref. [18], namely we estimate the kurtosis of log-returns for various lengths $T$ of windows and then select the optimal block width $T_{op}$ for which the average kurtosis $\bar{\kappa}_T$ over all windows is the

[2]In Refs. [47, 49] they used integration to improve estimation of the generalized Hurst exponent and thus they set $n = 1$.

[3]Note that log-returns $r$ are additive when changing to higher time scale.

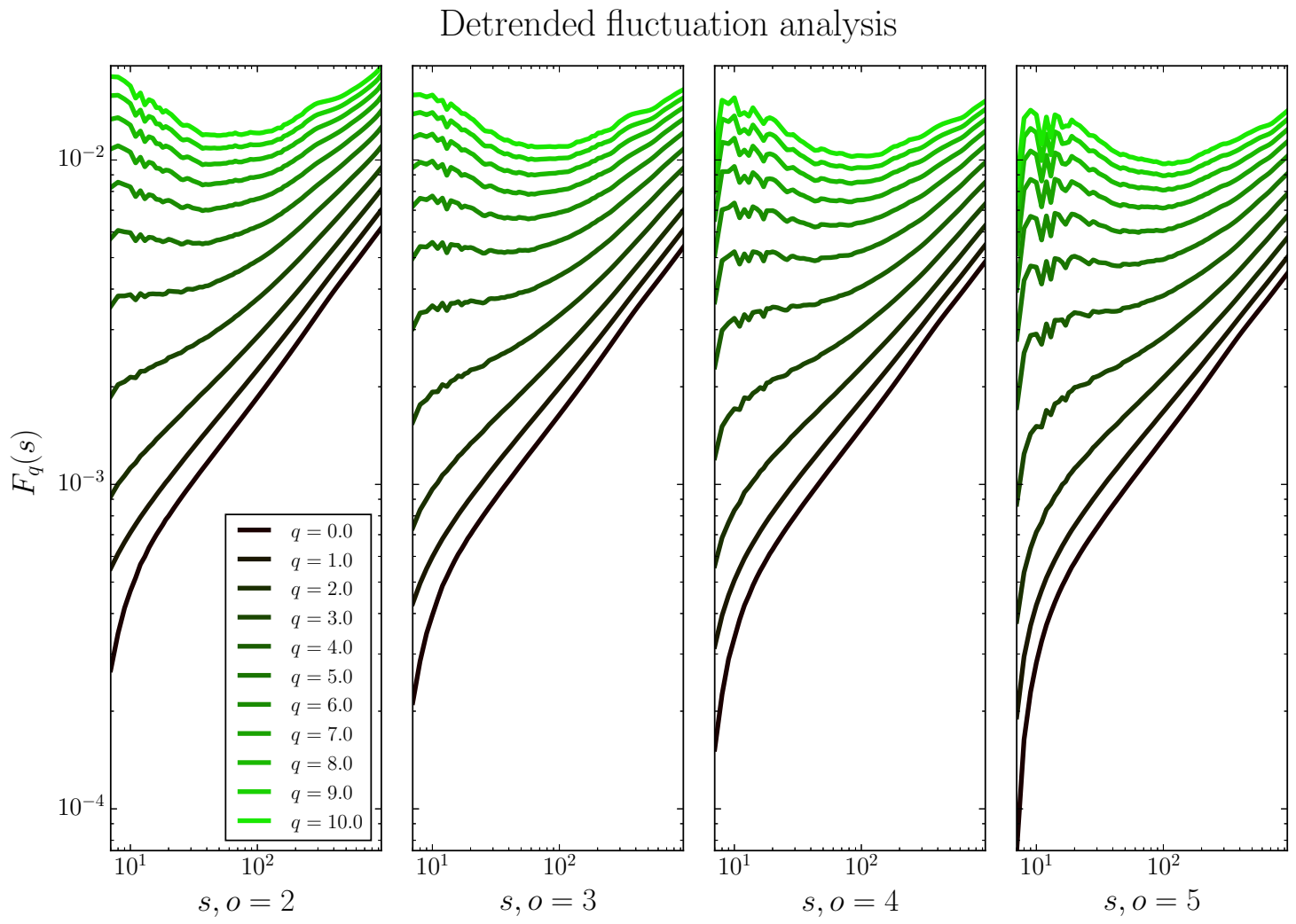

Figure 1. DFA analysis of log-returns of the AA company for 4 different orders of DFA. Apart from a short scale (less than 20 minutes), the behavior of $F_q(s)$ is quite robust. Depicted plots are on a log-log scale.

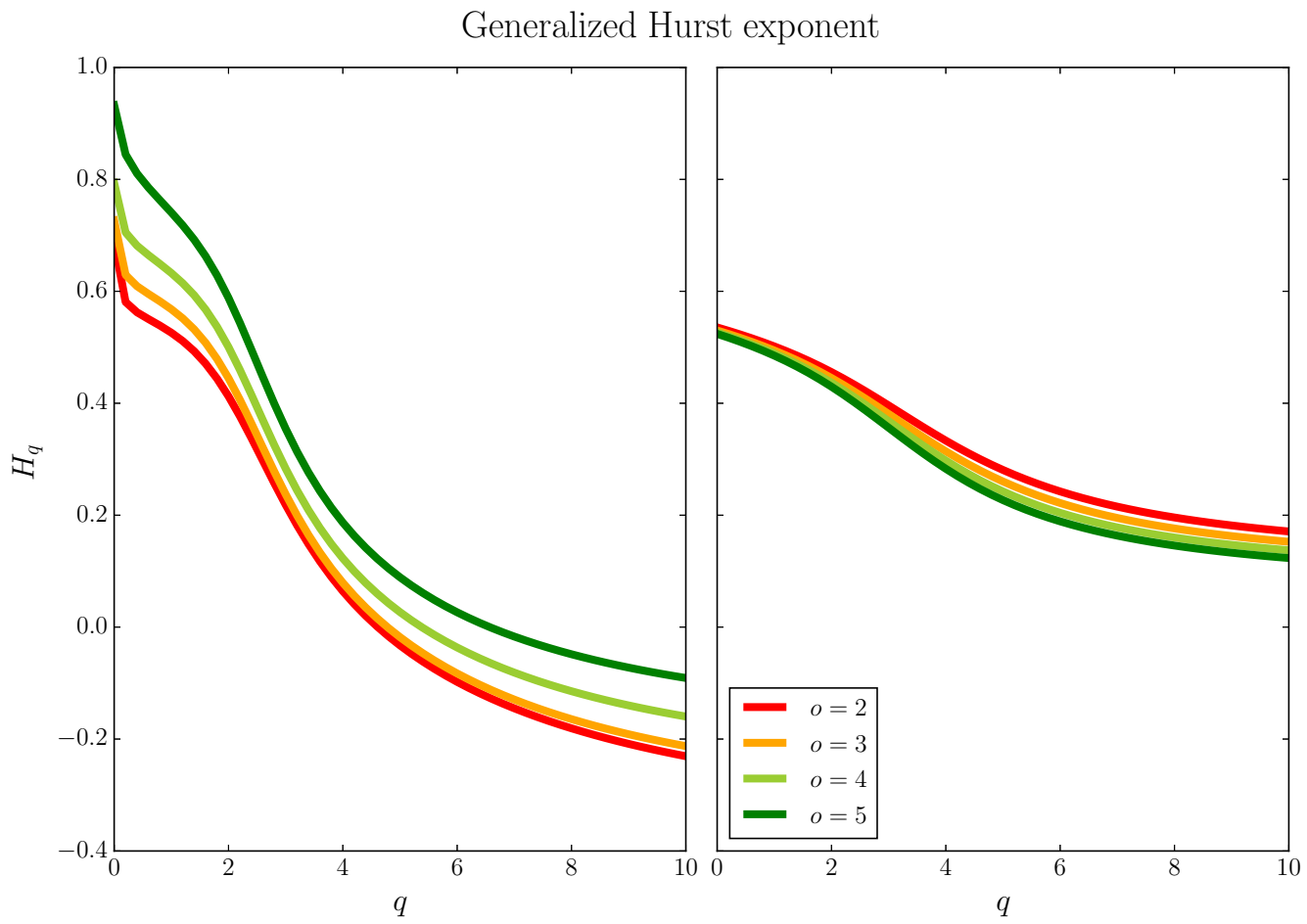

Figure 2. Generalized Hurst exponent as a function of $q$ for log-returns of the AA company in 2 different time scales. The respective time scale are $20 - 100$ minutes (left figure) and $150 - 500$ minutes (right figure).

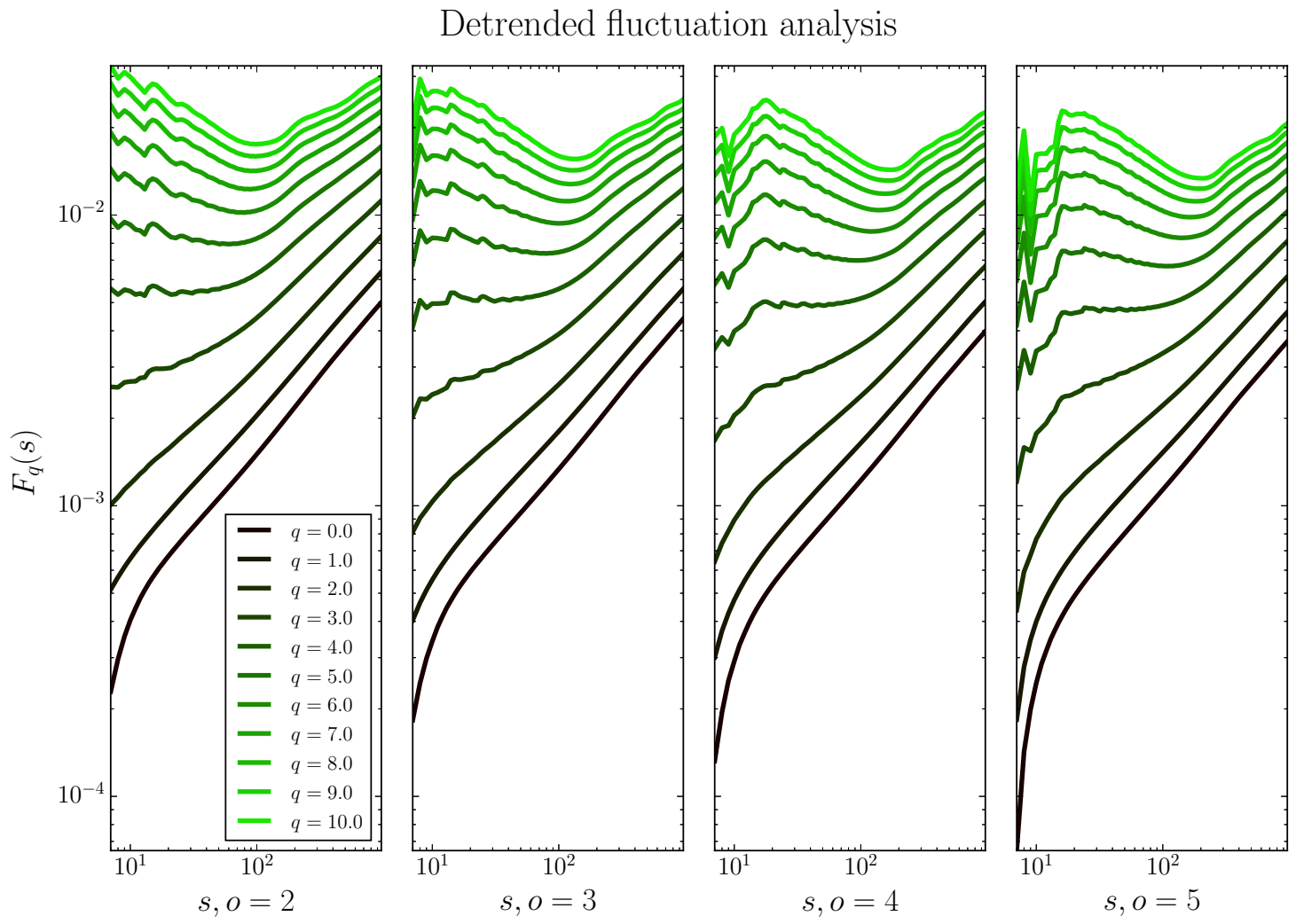

Figure 3. DFA analysis of log-returns of the BAC company for 4 different orders of DFA. Apart from a short scale (less than 20 minutes), the behavior of $F_q(s)$ is quite robust. Depicted plots are on a log-log scale.

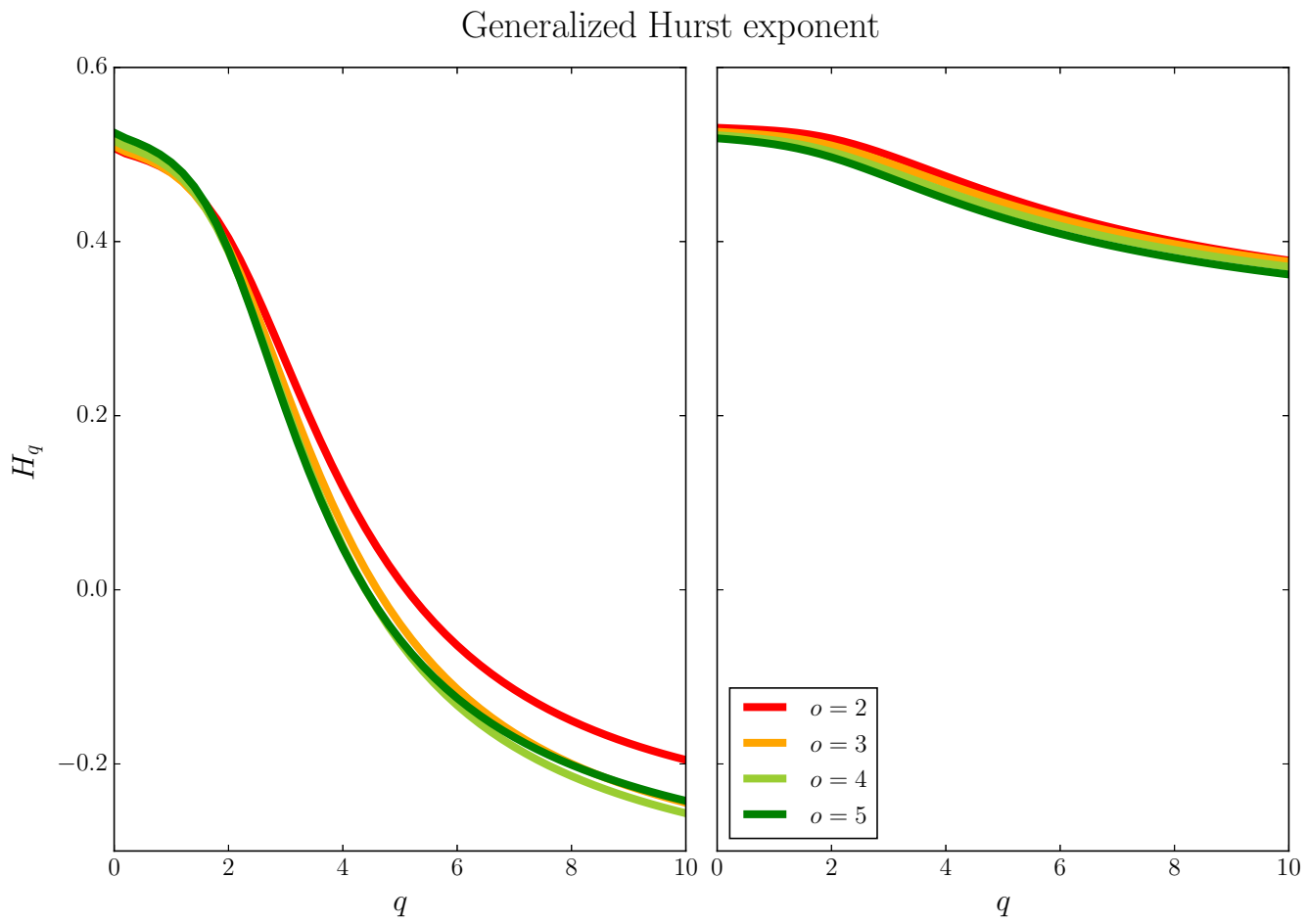

Figure 4. Generalized Hurst exponent as a function of $q$ for log-returns of the BAC company in 2 different time scales. The respective time scale are 20 − 100 minutes (left figure) and 150 − 500 minutes (right figure).

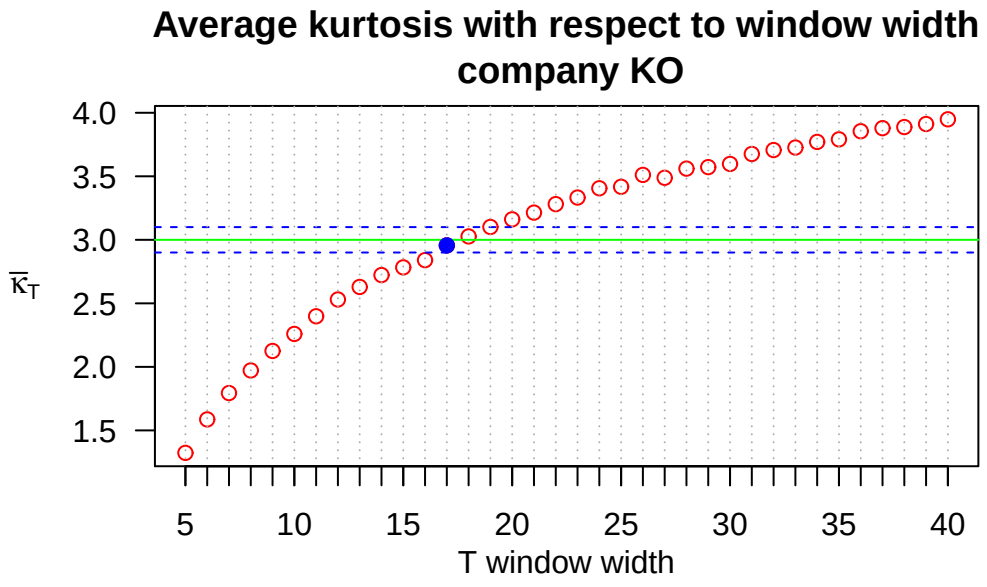

Figure 5. Finding optimal block width, $\epsilon = 0.1$.

closest to the one of the normal PDF, i.e., $\kappa = 3$. As a biased *moment estimator* for kurtosis we use

$$\hat{\kappa} = \frac{m_4}{m_2^2} = \frac{\frac{1}{T}\sum_{i=1}^{T}(r_i - \bar{r})^4}{\left[\frac{1}{T}\sum_{i=1}^{T}(r_i - \bar{r})^2\right]^2}, \tag{11}$$

where $\bar{r}$ is a sample mean of log-returns in a given time window. Apart from being a standard estimator, (11) has the lowest mean square error for normal sample, see, e.g., Ref. [38].

Since we are eventually interested in the distribution of variances, we prefer more data points for variance and the accuracy of estimated variance in each block is of a secondary importance because the errors cancel each other out in the density estimation. We introduce, therefore, a small threshold $\epsilon$ and consider the optimal $T$ as the lowest one for which the average kurtosis $\bar{\kappa}_T$ over $n = \lfloor N/T \rfloor$ blocks satisfies

$$|\bar{\kappa}_T - 3| < \epsilon . \tag{12}$$

We chose the threshold to be $\epsilon = 0.1$ which ensures the longest possible series of sample variances without significant departure from $\kappa = 3$, see Fig. 5.

For subsequent comparison of variance distributions on different scales it is convenient to normalize log-returns to zero mean and unit variance, i.e

$$u_i = \frac{r_i - \bar{r}}{s}, \tag{13}$$

where $\bar{r}$ and $s^2$ are sample mean and sample variance of the whole series, respectively. Having optimal window width and normalized log-returns, we estimate the variance in each block with the unbiased estimator

$$s_j^2 = \frac{1}{(T_{op} - 1)}\sum_{i=1}^{T_{op}}\left[u_{i+(j-1)T_{op}} - \mu_j\right]^2, \qquad j \in \left\{1, \dots, \lfloor N/T_{op} \rfloor\right\}. \tag{14}$$

Here $N$ is the length of the new series sampled every 20 min and $\mu_j$ denotes sample mean in a particular block

$$\mu_j = \frac{1}{T_{op}}\sum_{i=1}^{T_{op}} u_{i+(j-1)T_{op}} . \tag{15}$$

By this procedure we obtain $\lfloor N/T_{op} \rfloor$ values for variance (or "temperature" in physical jargon), however, in superstatistics we need the "inverse temperature" $\{\beta_i\}$ [c.f. Eq. (2)] which is

$$\beta_i = \frac{1}{s_i^2} . \tag{16}$$

In Fig. 6 the procedure for obtaining values of the inverse temperature is depicted.

Next step would be to find the mixing PDF for $\beta$. This task is in general quite intricate, but we can take advantage of the fact that in superstatistics there are only three two-parametric families of universal mixing distributions which are assumed to be relevant. These are the Gamma distribution, log-normal distribution and inverse Gamma distribution. Therefore, we face a much more tractable task of a parametric fitting where various methods are available, the most prominent being minimum distance method, moment method or maximum likelihood method. Here we confine ourselves to the maximum likelihood method in order not to interfere with distance measures used for goodness of fit.

When the optimal parameters are found, we can ask which of the three PDFs is the best fit for mixing PDF. Goodness of fit is usually measured by distances between expected distribution and empirical distribution. Here we employ three prototype distances, namely, Kolmogorov–Smirnov distance

$$D_n = \sup_x |F_n(x) - F(x)|, \tag{17}$$

Cramér–von Mises distance

$$C_n = n \int_{-\infty}^{+\infty} [F_n(x) - F(x)]^2 \mathrm{d}F(x), \tag{18}$$

and Anderson–Darling distance

$$A_n = n \int_{-\infty}^{+\infty} \frac{[F_n(x) - F(x)]^2}{F(x)[1 - F(x)]} \mathrm{d}F(x), \tag{19}$$

where $F(x)$ is fully specified expected distribution (i.e. all its parameters are given) and $F_n(x)$ is the empirical distribution

$$F_n(x) = \frac{1}{n} \sum_{i=1}^{n} I(u_i \leq x), \tag{20}$$

where $I(\cdots)$ is the indicator function. Corresponding statistical tests of goodness of fit make use of these distances in order to test the hypothesis that data come from a given distribution $F(x)$. Strictly speaking, a strong dependence of ACF on $\beta$ (see Fig. 7), makes the use of standard procedure and test of goodness of fit by these methods not warranted. Nevertheless, even for dependent data it makes sense to consider aforementioned distances as a convenient tool allowing to discriminate among the three PDFs. Our strategy therefore is to calculate distances for each company at a particular scale and choose the PDF with the smallest distance as the most optimal mixing PDF at that scale.

Since each distance measure has its own specific properties, it should not be surprising that different distances would yield different results. Yet our analysis shows that conclusions for all seven time series are quite robust and not very sensitive to the distance measures employed. In Tab. 1 we use the dataset of AA to illustrate the distance measures for the three mixing PDFs. Values correspond to scale 20 minutes which, as mentioned, is the smallest reliable scale. We can conclude from Tab. 1 (cf. also Fig. 8) that for small time scales the best superstatistics is the log-normal one. This is in agreement with the claim made in [18] where the statement was supported only by visual inspection of histogram and fitted distribution. In fact, the main point of [18] was to show that it is possible to observe different superstatistic at different time scales. The idea was demonstrated by fitting the PDF for $\beta$ at two remote time scales, namely, minute scale and daily scale. Tab. 2 supports the idea in a more quantitative way and we, indeed, see that a transition occurs, at least for company AA, from the log-normal distribution at small scales to the Gamma distribution at daily scale.

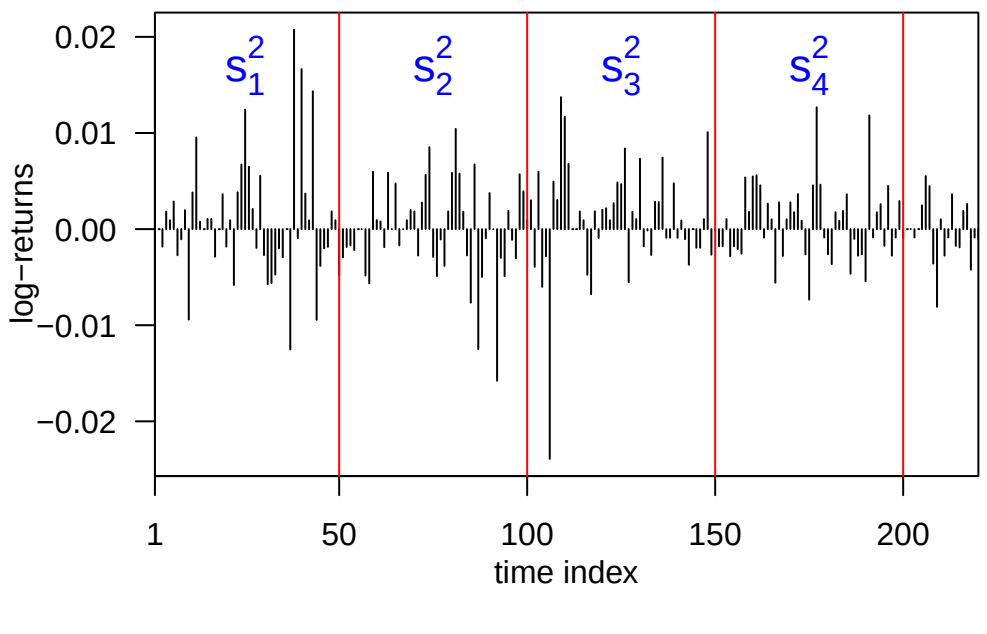

Figure 6. Illustrative figure for temperature estimation procedure.

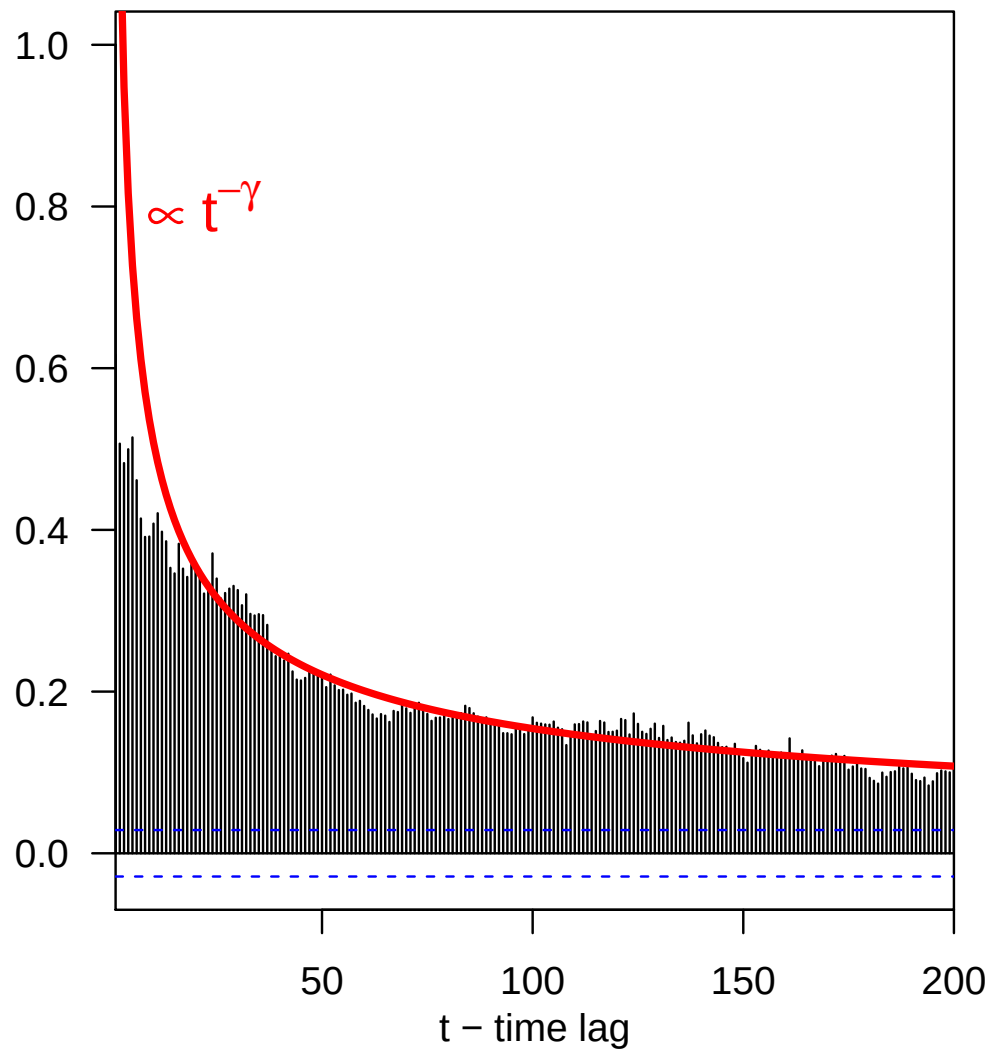

Figure 7. Autocorrelation function of $\beta$ for the AA company at scale 20 min. Best $\gamma$ fit provides $\gamma = 0.377$.

To get a better picture of the transition, it is convenient to calculate statistical distances for more than two time scales and see how the distances behave with respect to the time scale observed. We calculated distances for scales ranging from 20 minutes to 500 minutes $\sim$ 1 trading day (for higher scales more data would be needed). Results for short time scales of AA are depicted in Figs. 8 and 9 from which we see that only for very short times log-normal regime prevails. Unfortunately, this crude method does not allows to identify a *transition point* very reliably because when individual distance measures approach each other their respective characteristics should be taken into account and simple comparison of actual values cannot be considered as a decisive criterium. Nevertheless, in Fig. 10 we see that around 400 minutes we can reliably claim that change of superstatistics is already significant.

The above analysis together with results obtained quantitatively bolsters the observation reported in Ref. [18] for the AA company. However, according to [18] transition of superstatistics should appear in all seven analyzed companies. This is not the case (at least not for the observed time scales) as can be clearly seen from Figs. 11 and 12 (cf. also the Supplementary Material [50]) where the Anderson–Darling distance measure is shown for the company BAC (remaining two distance measures arrive to the same conclusion). Log-normal regime continues even on long time scales, and no transition of superstatistics is observed. This naturally does not exclude the existence of a transition on much longer time scales but clearly reveals drawback of a visual examination of the fitted histogram.

| | **Log-norm** | **Gamma** | **Inv-Gamma** |
|---|---|---|---|
| **Kolmogorov–Smirnov** | 0.032 | 0.050 | 0.117 |
| **Cramer–von Mises** | 1.26 | 3.04 | 22.06 |
| **Anderson–Darling** | 8.57 | 16.18 | 121.88 |

Table 1. Values for various distance measures for the three mixed PDFs, company AA, scale 20 min.

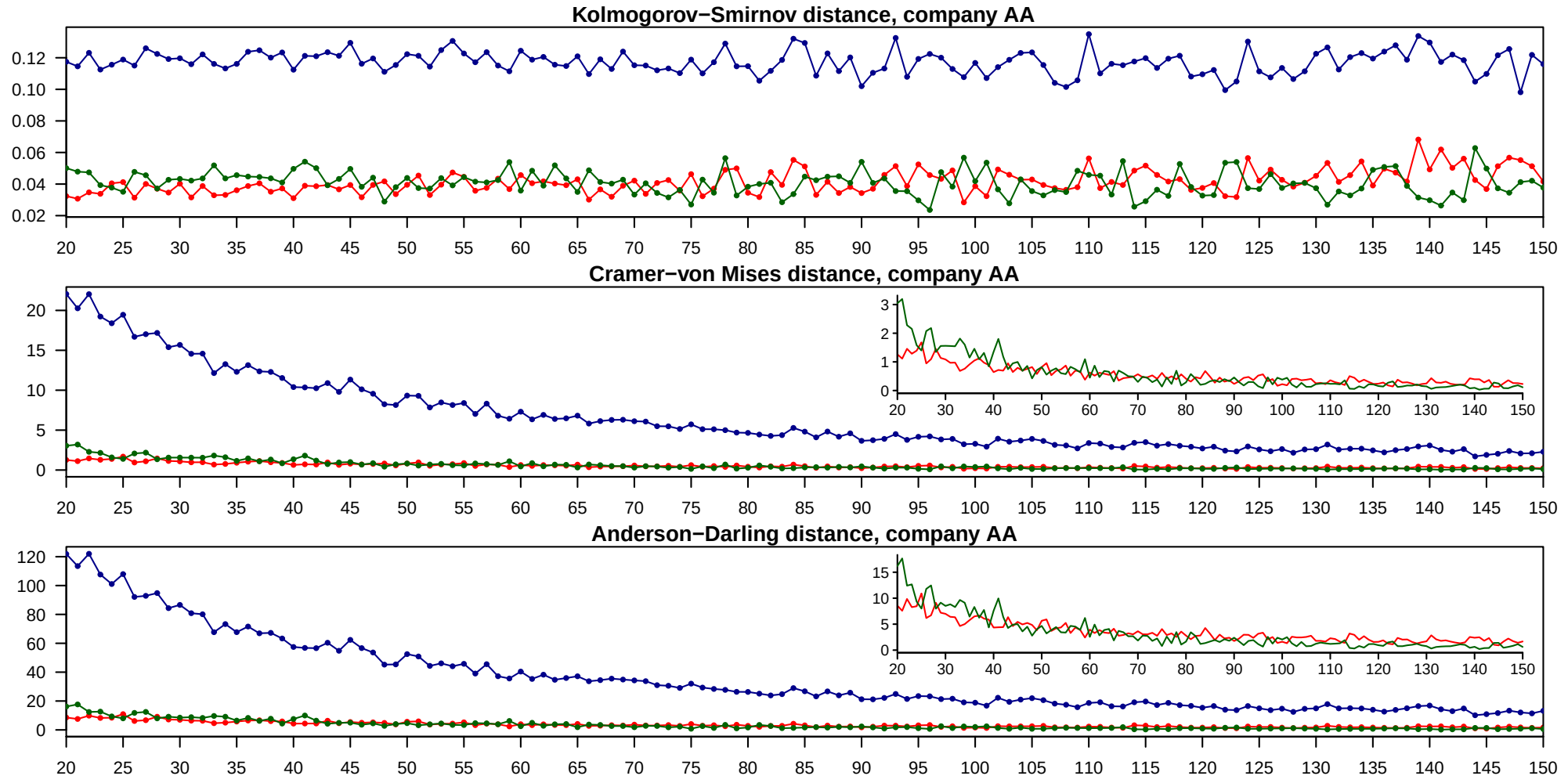

Figure 8. Dataset of AA: three statistical distance measures for considered superstatistics mixing PDFs are shown as a function of time scale in interval from 20 minutes to 2.5 hours (150 min). Blue: Inv-Gamma distribution, Green: Gamma distribution, Red: Log-normal distribution.

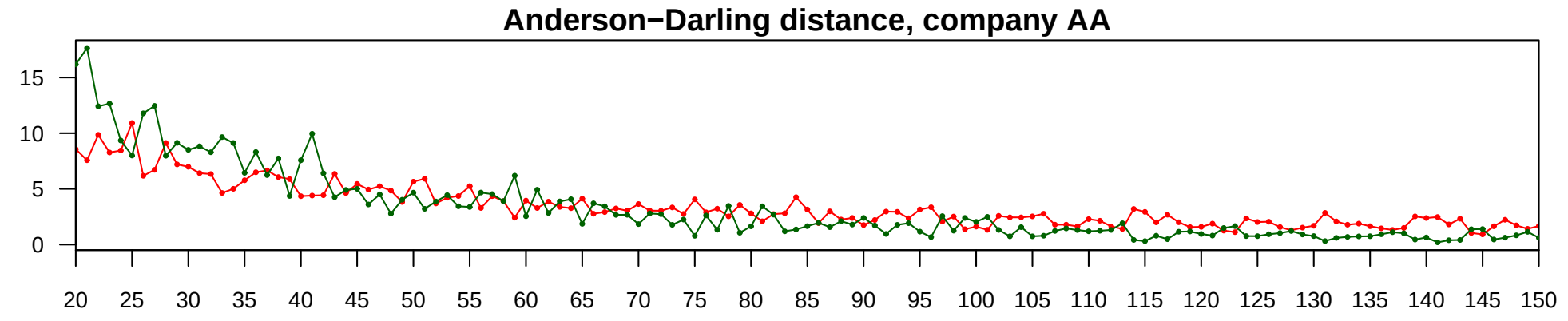

Figure 9. Detail of the Anderson–Darling distance measure for the AA company. The considered time scale is from 20 minutes to 2.5 hours (150 min). Green: Gamma distribution, Red: log-normal distribution. Inverse Gamma distribution is outlying and for simplicity not included.

In truth, authors of [18] admit that it is by no means easy to distinguish between different superstatistics, and this can be clearly understood from Figs. 10 and 12 where we see that the distance between log-normal and Gamma superstatistics are very close. It is also, in part, due to the fact that for higher-time scales we have less data points for "inverse temperature" $\beta$, and therefore statistical distances have less statistical power to distinguish between two probability distributions.

In total, we analyzed 7 companies (remaining 5 are reported in the Supplementary Material [50] out of which four, namely AA, INTC, KO and WMT, manifest the transition between superstatistics. On the other hand, companies BAC, GE and JNJ do not exhibit transition of superstatistics on the considered time scales. Interestingly, a very similar classification was obtained in Ref. [19], where the same time series were investigated with help of Renyi entropies in the symbolic space.

| | Log-normal | Gamma | Inv-Gamma |
|---|---|---|---|
| **Kolmogorov–Smirnov** | 0.077 | 0.045 | 0.133 |
| **Cramer–von Mises** | 0.153 | 0.032 | 0.770 |
| **Anderson–Darling** | 1.151 | 0.255 | 4.590 |

Table 2. Values of various distance measures for the three mixed PDFs, company AA, scale 390 min (roughly 1 trading day).

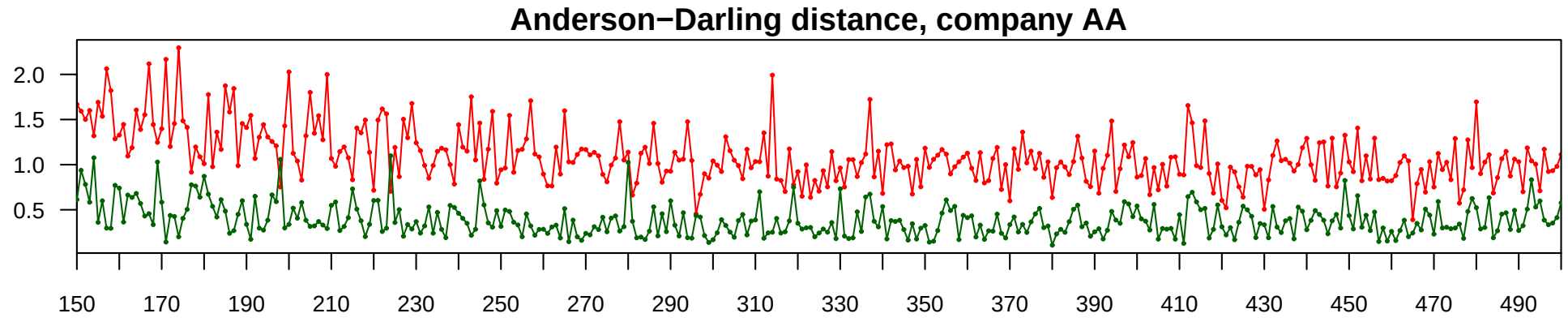

Figure 10. Detail of the Anderson–Darling distance measure for the AA company. The considered time scale is from from 2.5 hours (150 min) to 500 minutes (slightly over one trading day). Green: Gamma distribution, Red: log-normal distribution. Inverse Gamma distribution is outlying for simplicity not included.

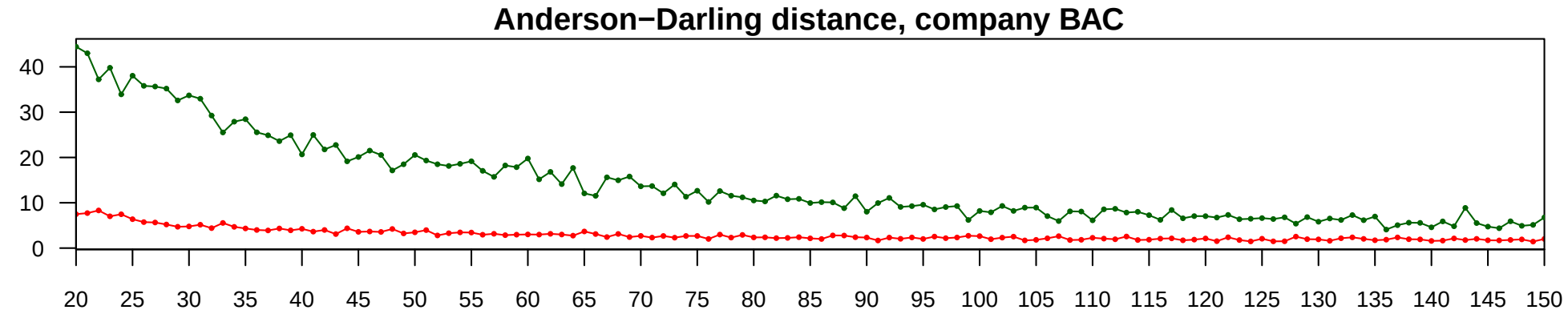

Figure 11. Dataset of BAC. Anderson–Darling distance measure for considered superstatistics mixing PDFs as a function of time scale in interval from 20 minutes to 2.5 hours (150 min). Green: Gamma distribution, Red: log-normal distribution. Inverse Gamma distribution is outlying and hence not included.

## 4. Regime switching between different superstatistical models: discussion

Previous numerical analysis illustrates that the transmutation of statistics observed in certain hierarchical complex systems can be understood as a transition between two superstatistics regimes. Apart from a relative simplicity of the numerical tests involved, this observation has also a heuristic value in that it allows one to cultivate some intuition about the underlying dynamics. To model the transition between two superstatistics, Xu and Beck proposed in Ref. [18], the so-called *synthetic model* where a single marginal variable $\beta$ fluctuates on a time scale that can accommodate the transition between log-normal superstatistics and Gamma-superstatistics. This is possible by defining $\beta$ as a random variable satisfying the linear interpolation ansatz

$$\beta \ = \ (1-\kappa)e^{X_0} \ + \ \kappa\sum_{i=1}^{n} X_i^2\,, \tag{21}$$

where $\kappa \in [0,1]$ is the interpolating (time-scale) parameter and all $X_i$ are independent identically distributed (IID) Gaussian variables. Also $X_0$ is a Gaussian variable but with generically different variance than $X_i$. One can expect that the dynamical variables $X_0$ and $X_i$ obey some underlying stochastic equations at their respective characteristic time scales. Some specific models of such stochastic dynamics were proposed in Ref. [18].

It is worthwhile to expand on this picture. Let us first consider a corresponding PDF for $\beta$ defined by Eq. (21). We will assume that $X_0$ takes values $x \in \mathbb{R}$, while $X_i$ values $x_i \in \mathbb{R}$. With this we have

$$g_\kappa(\beta) \ = \ \int_{\mathbb{R}^{n+1}} \mathrm{d}x_0 \mathrm{d}\mathbf{x}\, \delta\!\left(\beta - (1-\kappa)e^{x_0} - \kappa\sum_{i=1}^{n} x_i^2\right) \varphi_0(x_0)\varphi(\mathbf{x})\,, \tag{22}$$

where $\mathrm{d}\mathbf{x} = \prod_{i=1}^n \mathrm{d}x_i$ and $\varphi(\mathbf{x}) = \prod_{i=1}^n \varphi(x_i)$ is the characteristic function. Both $\varphi_0$ and $\varphi$ are Gaussian PDFs with zero mean and variances $\sigma_0^2$ and $\sigma^2$, respectively. The PDF associated with $\beta$ at the scale characteristic function associated

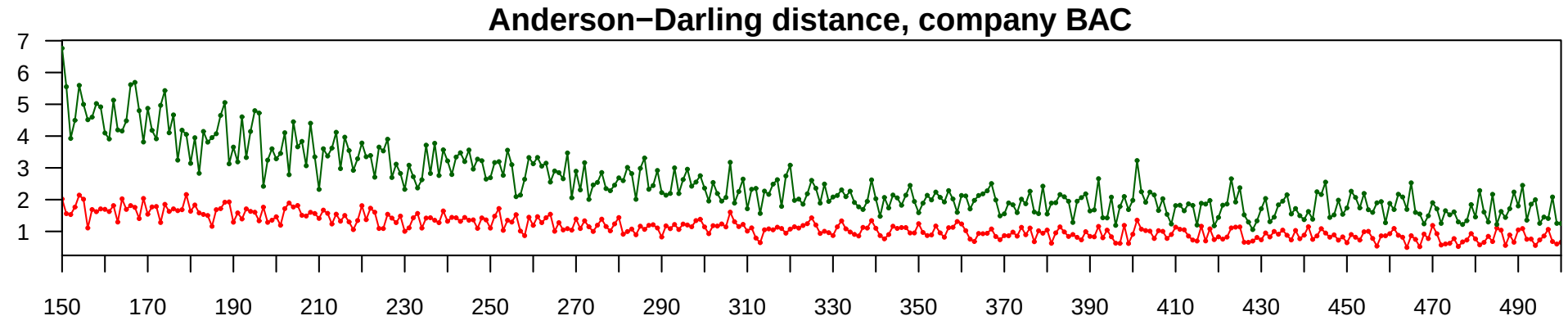

Figure 12. Dataset of BAC. Anderson–Darling distance measure for considered superstatistics mixing PDFs as a function of time scale in interval from 2.5 hours (150 min) to 500 minutes (slightly over one trading day). Green: Gamma distribution, Red: log-normal distribution. Inverse Gamma distribution is outlying and hence not included.

with $\beta$ has the form

$$
\begin{aligned}
\hat{g}_\kappa(t) &= \left[\int_{\mathbb{R}} dx_0\, \varphi_0(x_0)\, \exp\left[i(1-\kappa)te^{x_0}\right]\right]\left[\int_{\mathbb{R}} dx_1\, \varphi(x_1)\, e^{i\kappa t x_1^2}\right]^n \\
&= \left[\int_0^\infty \frac{dy_0}{y_0}\, \varphi_0(\ln y_0/(1-\kappa))\, e^{ity_0}\right]\left[\int_0^\infty \frac{dy_1}{\sqrt{y_1\kappa}}\, \varphi\left(\sqrt{y_1/\kappa}\right) e^{ity_1}\right]^n \\
&= \left[\int_0^\infty dy_0\, L\left(y_0; \ln(1-\kappa), \sigma_0\right)\, e^{ity_0}\right]\left[\int_0^\infty dy_1\, f(y_1; 1/2, 2\kappa\sigma^2)\, e^{ity_1}\right]^n \\
&= \left[\int_0^\infty dy_0\, L\left(y_0; \ln(1-\kappa), \sigma_0\right)\, e^{ity_0}\right]\left[\int_0^\infty dy_1\, f(y_1; n/2, 2\kappa\sigma^2)\, e^{ity_1}\right]. \qquad (23)
\end{aligned}
$$

Here $L(y_0; a, b)$ is the log-normal PDF with the *variance* $e^{2a+b^2}(e^{b^2}-1)$ and *mean* $e^{a+b^2/2}$, while $f(y_1; \alpha, \beta)$ is the Gamma PDF with the *variance* $\alpha\beta^2$ and *mean* $\alpha\beta$. The last line in (23) is a simple consequence of an infinite divisibility of the Gamma PDF. We see, in particular, that the number $n$ in (21) is responsible for the order of the mixing PDF in the Gamma superstatistics.[4] Clearly, when $\kappa \to 1$ log-normal distribution tends to the $\delta(y_0)$ and $\hat{g}_1(t)$ appraoches the characteristic function of the Gamma random variable $X$. Similarly, when $\kappa \to 0$ then $f(y_1; n/2, 2\kappa\sigma^2) \to \delta(y_1)$ and $\hat{g}_0(t)$ represents the characteristic function of the log-normal random variable. This is, of course, what one would expect from the convex combination of two random variables described via the *synthetic model* (21).

By employing the fact that both Gamma and log-normal random variables are infinitely divisible [27] (which is true also for their linear combination) we may rewrite the interpolation ansatz (21) in an equivalent, but for our purpose more expedient form, namely

$$
\beta = e^{X_0} + \frac{\ell}{m}\left(\sum_{i=1}^n X_i^2 - e^{X_0}\right) = X + Z = X + \sum_{i=1}^{\ell} Z_i\,. \qquad (24)
$$

Here $X$ is the log-normal random variable and $Z$ is a new infinite-divisible random variable with the characteristic function

$$
\hat{\mathcal{F}}_\ell(s) = \left[\int_0^\infty dy_0\, L\left(y_0; \ln(\ell/m), \sigma_0\right)\, e^{-isy_0}\right]\left[\int_0^\infty dy_1\, f(y_1; n/2, 2(\ell/m)\sigma^2)\, e^{isy_1}\right]. \qquad (25)
$$

In (24) the assumption was made that $\kappa = \ell/m$ is a rational number with $\ell, m \in \mathbb{N}$ and $0 \leq \ell \leq m$. In particular, for $\ell = 0$ we set $\sum_{i=1}^0 Z_i = 0$. Any random variable $Z_i$ has thus the characteristic function $[\hat{\mathcal{F}}_\ell(s)]^{1/\ell}$. This can be inverted to obtain the PDF $\mathcal{G}_{\ell/m}(z_i)$ for $Z_i$. Unfortunately, the latter cannot be done in a closed analytical form as only approximative or numerical approaches are available.[5] Note that the RHS of (24) describes the initial (short-time-scale) log-normal random variable to which we add at equidistant time intervals $1/m$ equally distributed random variables $Z_i$. This implies that the statistical influences which determine the form of $\beta$ are at given time instances $\ell/m$, $0 \leq \ell \leq m$, identical. The actual form of the distribution $\mathcal{G}_{\ell/m}(z)$ depends at each instance on the size $1/m$ of the time interval considered. After $\ell$ additions this brings us to the mixed random variable $\beta$ at time $\kappa = \ell/m$. The associated PDF has according to (24) the form

$$g_{\ell/m}(\beta) \ = \ \int \mathrm{d}x \mathrm{d}\mathbf{z}\, \delta\left(\beta - x - \sum_{i=1}^{\ell} z_i\right) L(x; 0, \sigma_0) \prod_{i=1}^{\ell} \mathcal{G}_{\ell/m}(z_i)\,. \tag{26}$$

[4] Equivalently, one can view $n$ as the number of degrees of freedom in the $\chi^2$ superstatistics.

At this stage it is convenient to pass to new variables $\beta_i \in [0, \infty)$ via transformation:

$$\begin{aligned} x &= \beta_1 - \beta_0\,, \\ z_1 &= \beta_2 - \beta_1\,, \\ z_2 &= \beta_3 - \beta_2\,, \\ &\vdots \\ z_{\ell-1} &= \beta_\ell - \beta_{\ell-1}\,, \\ z_\ell &= \beta_{\ell+1} - \beta_\ell\,. \end{aligned} \tag{27}$$

Here $\beta_0 = 0$. Transformation (27) converts a stationary time series $\{z_j\}_{j=1}^{\ell}$ to a random walk (Lévy diffusion trajectory) where the value of a sample path at time $\kappa = \ell/m$ is $\beta_{\ell+1} = \beta_1 + \sum_{j=1}^{\ell} z_j$. With (27) we can recast (26) into

$$g_{\ell/m}(\beta) \ = \ \int_0^\infty \mathrm{d}\beta_1 \mathrm{d}\beta_2 \ldots \mathrm{d}\beta_{\ell+1}\; \delta(\beta - \beta_{\ell+1})\; L(\beta_1; 0, \sigma_0) \prod_{i=1}^{\ell} \mathcal{G}_{\ell/m}(\beta_{i+1} - \beta_i)\,, \tag{28}$$

or more explicitly

$$\begin{aligned} g_0(\beta) &= L(\beta; 0, \sigma_0)\,, \\ g_{\ell/m}(\beta) &= \int_0^\infty \mathrm{d}\beta_1 \mathrm{d}\beta_2 \ldots \mathrm{d}\beta_\ell\; \mathcal{G}_{\ell/m}(\beta - \beta_\ell) \prod_{i=1}^{\ell-1} \mathcal{G}_{\ell/m}(\beta_{i+1} - \beta_i)\; L(\beta_1; 0, \sigma_0)\,, \end{aligned} \tag{29}$$

for $\ell \neq m$, and

$$g_1(\beta) \ = \ \int_0^\infty \mathrm{d}\beta_1 \mathrm{d}\beta_2 \ldots \mathrm{d}\beta_m\; \mathcal{G}_1(\beta - \beta_m) \prod_{i=1}^{m-1} \mathcal{G}_1(\beta_{i+1} - \beta_i)\; L(\beta_1; 0, \sigma_0) \ = \ f(\beta; n/2, 2m\sigma^2)\,, \tag{30}$$

for $\ell = m$. Note that $g_1(\beta)$, by its very construction, is the Gamma distribution $f(\beta; n/2, 2m\sigma^2)$. In particular from (29) and (30), we see that

$$\prod_{i=1}^{\ell} \mathcal{G}_{\ell/m}(\beta_{i+1} - \beta_i)\; L(\beta_1; 0, \sigma_0) \ \equiv \ \wp(\beta_1, \beta_2, \ldots, \beta_{\ell+1})\,, \tag{31}$$

represents the joint PDF of $\beta_1, \ldots, \beta_{\ell+1}$. For the marginal PDF at time scale $\kappa = \ell/m$ we may thus write

$$\begin{aligned} \wp_\ell(A) &= \int_0^\infty \mathrm{d}\beta_1 \ldots \mathrm{d}\beta_{\ell+1}\, \wp_\ell(A|\beta_1, \beta_2, \ldots, \beta_{\ell+1})\, \wp(\beta_1, \beta_2, \ldots, \beta_{\ell+1}) \\ &= \int_0^\infty \mathrm{d}\beta_1 \ldots \mathrm{d}\beta_{\ell+1}\, \wp_\ell(A|\beta_1, \beta_2, \ldots, \beta_{\ell+1})\, \wp(\beta_1, \beta_2, \ldots, \beta_\ell | \beta_{\ell+1})\, g_{\ell/m}(\beta_{\ell+1}) \\ &= \int_0^\infty \mathrm{d}\beta\, \wp_\ell(A|\beta)\, g_{\ell/m}(\beta)\,, \end{aligned} \tag{32}$$

[5] A closed-form formula for the characteristic function of the log-normal PDF is not analytically known and only approximative formula in terms of Lambert-W functions is available [27, 29].

where on the last line we have defined the averaged conditional PDF over fast variables $\{\beta_i\}_{i=1}^{\ell}$:

$$\wp_\ell(A|\beta) \ = \ \int_0^\infty d\beta_1 \ldots d\beta_\ell \, \wp(A,\beta_1,\beta_2,\ldots,\beta_\ell|\beta_{\ell+1})\,. \tag{33}$$

Note that the last equality in (32) corresponds to the canonical superstatistics at the time scale $s_\ell$, provided that $\wp_\ell(A|\beta)$ is a *bona fide* conditional probability. By construction, this is the case here only for $\ell = 0$ and $\ell = m$.

Let us now see that the above reformulation of the synthetic model with rationally-valued interpolation time $\kappa$ can be naturally embedded within a multi-scale superstatistics framework (4). This is turn will provide a simple stochastic picture connecting the two asymptotic "canonical" superstatistics. To this end we will consider a sequence of scales $\{s_i = s_0 + i/m; i = 0,\ldots,\ell\}$ which are characteristic for $\{\beta_{i+1}\}_{i=0}^{\ell}$. At the shortest scale $s_0 \sim 50$ minutes (described by the nuisance parameter $\beta_1$) the observed superstatistics is described by the relation

$$\wp_0(A) \ = \ \int_0^\infty d\beta_1 \, \wp(A|\beta_1, \text{s.v.}) \, \wp_1(\beta_1|\text{s.v.})\,. \tag{34}$$

Here the abbreviation "s.v." denotes the variables slower than $\beta_1$. The smearing PDF $\wp_1(\beta_1|\text{s.v.})$ is in our case the log-normal PDF. On the other hand, at an intermediate scale where $\kappa = \ell/m < 1$ (i.e., when $s_\ell \ll s_m \sim 500$ minutes) we can start with the joint PDF $\wp(A,\beta_1,\beta_2,\ldots,\beta_{m+1})$ and integrate over the shorter time scales shorter than $s_\ell$ (which is equivalent to coarse-graining or averaging over fast variables). This leads to

$$\wp_\ell(A,\beta_{\ell+1}, \text{s.v.}) \ = \ \int_0^\infty d\beta_1 \ldots d\beta_\ell \, \wp(A,\beta_1,\beta_2,\ldots,\beta_{m+1}) \ = \ \wp_\ell(A|\beta_{\ell+1}, \text{s.v.}) \wp_{\ell+1}(\beta_{\ell+1}, \text{s.v.})\,. \tag{35}$$

The last equality in (35) results from the De Finetti–Kolmogorov relation. Acronym, "s.v." denotes the variables at time scales larger than $s_\ell$, i.e., slow[6] variables $\{\beta_{i+1}\}_{i=\ell+1}^{m}$. So, when one concentrates on the behavior at the scale $s_\ell$, "s.v." appear only as external parameters in the PDF. The corresponding marginal PDF reads

$$\begin{aligned}
\wp_\ell(A) \ \equiv \ \wp_\ell(A, \text{s.v.}) \ &= \ \int_0^\infty d\beta_{\ell+1} \, \wp_\ell(A|\beta_{\ell+1}, \text{s.v.}) \, \wp_{\ell+1}(\beta_{\ell+1}, \text{s.v.}) \\
&= \ \int_0^\infty d\beta_1 \ldots d\beta_{\ell+1} \prod_{i=1}^{\ell} \wp_i(\beta_i|\beta_{i+1},\beta_{i+2},\ldots\beta_{\ell+1}, \text{s.v.}) \, \wp(A|\beta_1,\beta_2,\ldots,\beta_{\ell+1}, \text{s.v.})\,.
\end{aligned} \tag{36}$$

On the last line we have used Eq. (3) and (35). Here, "s.v." denotes the variables slower than $\beta_{\ell+1}$, i.e., $\{\beta_{i+1}\}_{i=\ell+1}^{m}$. So, when one concentrates on the behavior at the scale $s_\ell$, "s.v." appear only as external parameters in the PDF.

At the largest characteristic scale $\sim 500$ minutes (i.e., when $\ell = m$ or equivalently $\kappa = 1$) we have

$$\begin{aligned}
\wp_m(A) \ \equiv \ \wp(A) \ &= \ \int_0^\infty d\beta_1 d\beta_2 \ldots d\beta_{m+1} \, f(\beta_{m+1}) \prod_{i=1}^{m} \wp_i(\beta_i|\beta_{i+1},\beta_{i+2},\ldots,\beta_{m+1}) \, \wp(A|\beta_1,\beta_2,\ldots,\beta_{m+1}) \\
&= \ \int_0^\infty d\beta_{m+1} \, \wp_0(\beta_{m+1}) \, \wp(A|\beta_{m+1})\,,
\end{aligned} \tag{37}$$

where in the second equality we have employed Eq. (3) and the marginalization over $\beta_1,\ldots,\beta_m$, namely

$$\int_0^\infty d\beta_1 \ldots d\beta_m \ \wp(A,\beta_1,\beta_2,\ldots,\beta_{m+1}) \ = \ \wp(A,\beta_{m+1}) \ = \ \wp(A|\beta_{m+1}) \wp_0(\beta_{m+1})\,. \tag{38}$$

The last equality in (38) is again a consequence of the De Finetti–Kolmogorov relation. For share-price returns at hand the empirical evidence indicates that at the scale $s_m$ the marginal PDF follows a "canonical" form of the Gamma superstatistics which allows to identify the smearing PDF $\wp_0(\beta_{m+1})$ with the Gamma PDF.

[6]The slowness at the time scales $s_\ell$ can be quantified by a typical spread of the array of sample paths $\beta_t$ in the Lévy diffusion process at the time instant $s_\ell$. As a measure of the spread one can use, e.g., local variance or an appropriate local fractional moment.

By comparing (29)-(30) with (34)-(37) we can make the following identifications:

$$
\begin{aligned}
\wp_\ell(A|\beta_{\ell+1})g_{\ell/m}(\beta_{\ell+1}) &= \wp_\ell(A|\beta_{\ell+1},\text{s.v.})\wp_{\ell+1}(\beta_{\ell+1},\text{s.v.})\,,\\
\wp_\ell(A|\beta_1,\beta_2,\ldots,\beta_{\ell+1}) &= \wp(A|\beta_1,\beta_2,\ldots,\beta_{\ell+1},\text{s.v.})\,,\\
\mathcal{G}_{\ell/m}(\beta_{i+1}-\beta_i) &= \wp_i(\beta_i|\beta_{i+1},\beta_{i+2},\ldots\beta_\ell,\text{s.v.})\,,\quad i>1\,,\\
L(\beta_1;0,\sigma_0) &= g_0(\beta_1) = \wp_1(\beta_1|\text{s.v.})\,,\\
f(\beta_{m+1};n/2,2m\sigma^2) &= g_1(\beta_{m+1}) = \wp_0(\beta_{m+1})\,. \qquad (39)
\end{aligned}
$$

Notice in particular that the synthetic model model can be identified with the multi-scale superstatistics in the Markovian approximation (4).

Let us finally comment on Eq. (32). There the *effective* conditional PDF $\wp_\ell(A|\beta)$ was defined by integrating over fast variables $\{\beta_i\}_{i=1}^{\ell}$, i.e., up to time scale $s_{\ell+1}$ [cf. Eq. (33)]. In fact, only for $\ell = 0$ and $\ell = m$ the *effective* conditional PDF turns out to be a genuine conditional PDF and only in these cases (32) represents a true canonical superstatistics. This is akin to the renormalization group (RG) approach [31] where close to the critical point the correlation length is the only important (length) scale, and that the microscopic (length) scales are irrelevant. In our case $\wp_\ell(A|\beta)$ follows a very similar pattern of behavior, namely at the scales where the canonical superstatistics appear only one scale is relevant (namely $\beta_1$ or $\beta_{m+1}$) and all other scales are *irrelevant.* [7] The scales at which the canonical superstatistics appear can thus be understand as critical points — stable fixed points. In addition, the interpolation formula (32) between our two critical points can be understand as a RG flow equation. Work along these lines is presently under investigation.

## 5. Breakdown of superstatistics: threshold of validity

While in the previous two sections we have seen that superstatistics might serve as a useful device allowing to indicate and analyze transmutation of statistics on various time scales, in the present section we will briefly discuss some of the danger involved an uncritical use of thereof.

Notably, in many complex systems it is possible to express some of relevant PDFs in a form that bears a close resemblance to the superstatistical marginal distribution (1). Nevertheless, such representations are, as a rule, only formal and even though they may have some heuristic value, they sometimes lead to false empirical signatures which in turn indicate a breakdown of an unwarranted superstatistical picture. In the following we will illustrate this point with two examples.

### 5.1. Example 1: superstatistical interpretation of the space-time fractional diffusion

Here we will concentrate on space-time fractional diffusion processes. These processes are based on generalized (Fokker–Planck-type) diffusion equation, where ordinary derivatives are replace by so-called *fractional derivatives*. Details on the space-time fractional diffusion and its applications in finance are discussed, e.g., in Refs. [32, 33, 34]. It is usually written the form [35]

$$\left({}_0^*\mathcal{D}_t^\gamma - {}^\theta\mathcal{D}_x^\alpha\right)\wp(x,t) = 0\,, \qquad (40)$$

where

$${}_{t_0}^{*}\mathcal{D}_t^\gamma f(t) = \frac{1}{\Gamma(\lceil\gamma\rceil-\gamma)}\int_{t_0}^t \frac{f^{\lceil\gamma\rceil}(\tau)}{(t-\tau)^{\gamma+1-\lceil\gamma\rceil}}\,\mathrm{d}\tau\,, \qquad (41)$$

($\lceil\cdots\rceil$ represents the ceiling function) is the so-called Caputo fractional derivative and ${}^\theta\mathcal{D}_x^\alpha$ is the so-called (generalized) Riesz fractional derivative which is a pseudo-differential operator defined via the Fourier transform [8]

$$\mathcal{F}[{}^\theta\mathcal{D}_x^\alpha f(x)](k) = -{}^\theta\psi^\alpha(k)\mathcal{F}[f(x)](k) = -|k|^\alpha \exp\left[i\,\text{sign}(k)\,\theta\pi/2\right]\mathcal{F}[f(x)](k)\,. \qquad (42)$$

---

[7] Here mean that as the RG flow is attracted to the fixed point the *irrelevant* degrees of freedom can be neglected.

[8] Conventional Riesz fractional derivative has $\theta = \pm 1$.

Interestingly, the solution of the space-fractional diffusion equation with pseudo-differential operator ${}^{\theta}\mathcal{D}_x^{\alpha}$ leads to $\alpha$-stable diffusion $L_{\alpha,\theta}(x,t)$. The parameters have to satisfy $0 < \gamma \leq \alpha \leq 2$ in order to get the probabilistic interpretation of $\wp$ (i.e., $\wp$ is a positive function defined on the $L^1(\mathbb{R}, \mathrm{d}x)$ space).

The solution of Eq. (40) can be found through the Laplace–Fourier transform ($t \overset{\mathcal{L}}{\leftrightarrow} s$, $x \overset{\mathcal{F}}{\leftrightarrow} k$):

$$\hat{\bar{\wp}}^{\theta}_{\alpha,\gamma}(k,s)\, s^{\gamma} - s^{\gamma-1} + {}^{\theta}\psi^{\alpha}(k)\hat{\bar{\wp}}^{\theta}_{\alpha,\gamma}(k,s) \;=\; 0\,. \tag{43}$$

This trivially gives

$$\hat{\bar{\wp}}^{\theta}_{\alpha,\gamma}(k,s) \;=\; \frac{s^{\gamma-1}}{s^{\gamma} + {}^{\theta}\psi^{\alpha}(k)}\,. \tag{44}$$

The solution can be found by applying the Mellin–Barnes integral transform [35]. Let us concentrate on a different representation of $\wp^{\theta}_{\alpha,\gamma}$. With help of Schwinger's trick [15] we can rewrite (44) in the form

$$\hat{\bar{\wp}}^{\theta}_{\alpha,\gamma}(k,s) \;=\; \int_0^{\infty} \mathrm{d}l\, s^{\gamma-1} e^{-ls^{\gamma}} e^{-l^{\theta}\psi^{\alpha}(k)} \;=\; \int_0^{\infty} \mathrm{d}l\; \hat{\wp}_{\gamma}(l,s)\bar{\wp}^{\theta}_{\alpha}(k,l)\,. \tag{45}$$

After the inverse Fourier–Laplace transform we end up with

$$\wp^{\theta}_{\alpha,\gamma}(x,t) \;=\; \int_0^{\infty} \mathrm{d}l\, \wp_{\gamma}(l,t)\wp^{\theta}_{\alpha}(x,l)\,. \tag{46}$$

Here $\wp^{\theta}_{\alpha}(x,l)$ is the $\alpha$-stable distribution and $\wp_{\gamma}(l,t)$ can be considered as a smearing kernel, where $l$ can be viewed as the nuisance parameter (basically pseudo-time).

Interestingly, both $\wp^{\theta}_{\alpha}(x,l)$ and $\wp_{\gamma}(l,t)$ are solutions of (single)-fractional equations

$$\frac{\mathrm{d}\wp^{\theta}_{\alpha}(x,l)}{\mathrm{d}l} \;=\; {}^{\theta}\mathcal{D}^{\alpha}_{x}\wp^{\theta}_{\alpha}(x,l)\,, \tag{47}$$

$$\frac{\mathrm{d}\wp_{\gamma}(l,t)}{\mathrm{d}l} \;=\; \mathcal{D}^{\gamma}_{t}\wp_{\gamma}(l,t)\,, \tag{48}$$

and the resulting $\wp^{\theta}_{\alpha,\gamma}(x,t)$ is given by Eq. (46). The Schwinger integral representation formally corresponds to superstatistical formulation in Eq. (4). On the other hand, there are several aspects which need to be discussed before we conclude that a given system can be treated as a genuine superstatistical system:

- *Probabilistic interpretation*: in order to give a clear interpretation to the smearing distribution $\wp_{\gamma}(l,t)$, it must be a positive normalized function. This is true only for $\gamma < 1$. For $\gamma > 1$, $\wp_{\gamma}(l,t)$ cannot be interpreted as the superstatistical smearing kernel even if the time scales are well separated. Let us note that this is true only for one-dimensional diffusion. For multivariate case the situation is even more complex, see Refs. [36, 37].

- *Separated time scales:* Even for $\gamma < 1$ it is not automatically guaranteed that the solution results from superstatistics. For superstatistics to work it is necessary to have two well separated time-scales. For short-time scales, the solution can be sufficiently well described by the $\alpha$-stable distribution $\wp^{\theta}_{\alpha}(x,t)$, while for long time scales, we have to use the smeared distribution $\wp^{\theta}_{\alpha,\gamma}(x,t)$.

As a result, a small shift of parameters in Eq. (46) can lead to a superstatistics transition or even breakdown. This can be illustrated on the model of space-time fractional diffusion of varying order described in Ref. [34]. This simple time-dependent model considers distinct time intervals $T_i = [t_{t-1}, t_i]$. In each interval the system is described by the space-time double fractional diffusion with parameters $(\alpha_i, \theta_i, \omega_i)$. Let us further assume that $\theta_i = \theta$ is constant (typically $\theta = 0$ in physical applications [32] or $\theta = -1$ in financial applications [33, 34]) and $\gamma_i/\alpha_i = \Omega \in (0,1]$ is also constant describing scaling of the distribution, so $\wp(x,t) = t^{-\Omega}F\left(xt^{-\Omega}\right)$. Thus, the distribution in every interval is described by a single parameter $\alpha_i$. Let us assume, for simplicity, that $|T_i| = \tau$ is constant. Then for large times $t \gg \tau$ it is possible to assume that particular choice of parameter $\alpha_i$ is described by a random variable with the PDF $f(\alpha)$. The resulting distributions for long times can be understood as a 3-scale superstatistics

$$\wp^{\theta}_{\Omega}(x,t) \;=\; \int_0^{2}\mathrm{d}\alpha\int_0^{\infty}\mathrm{d}l\, f(\alpha)\, \wp_{\alpha\Omega}(l,t)\, \wp^{\theta}_{\alpha}(x,l)\,. \tag{49}$$

Naturally, the validity of such description always depends on a particular system at hand. Here we have to add, apart from previous points, the assumption on different scales and their separability. Actually, for times $t \ll \tau$, the system is described by the $\alpha$-stable distribution; for $t \approx \tau$ the evolution is given by a space-time fractional diffusion on a given interval $T_i$ and for $t \gg \tau$ we have yet another smearing over $\alpha$.

### 5.2. Example 2: Brownian subordination

Another important example where one can encounter a potential breakdown of the canonical superstatistics occurs in connection with Brownian subordination in the theory of Lévy processes. The connection of superstatistics to Brownian subordinators was recently discussed e.g. in Ref. [41].

Let us recall that Lévy processes are processes with stationary and independent increments. A central result in the theory of Lévy processes is the so-called Lévy–Itô decomposition [27], which states that any Lévy process of bounded variance can be written as a sum of a Brownian component, a linear drift (trend component) and a pure jump Poisson-type process. Statistical properties of the pure jump process are fully described by Lévy measure $\nu$, which is defined for general Lévy processes $X_t$ as

$$\nu(A) \ = \ \mathbb{E}(\nu_t(A))|_{t=1}\,, \tag{50}$$

where the random measure $\nu_t$ counts the "jumps" of $X$ of size $A$ occurring within the time interval $(0, t]$, i.e.

$$\nu_t(A) \ = \ \sum_{0<s\leq t} I(\Delta X_s \in A, \Delta X_s \neq 0)\,. \tag{51}$$

Here $\Delta X_t = X(t) - X(t_-)$. Accordingly, the measure $\nu$ represents an average number of jumps per unite time.

An important subclass of Lévy processes are *subordinators* which consist of processes with (strictly) increasing trajectories. Subordinators have no diffusion component, only a positive drift and positive jump sizes, which thus implies that $\nu(\mathbb{R}^-) = 0$. Subordinator can be used as a new stochastic time, in which case one speaks about *subordination*. Conceptual roots of subordination can be traced back to a series of seminal papers of Mandelbrot et al. [51, 52] on a multifractal model of asset returns. When the original process is a Brownian motion then the ensuing subordination is known as a Brownian subordination.

Let us consider a stochastic process $B_t = \sigma W_t + \mu t$, where $W_t$ is standard Wiener process. Further we will consider a subordinator $T_t$ with a drift $c$ and associated absolutely continuous Lévy measure $\rho(t)\mathrm{d}t$. We can then use a Brownian subordination to define a new process

$$X_t \ = \ \sigma W_{T_t} \ + \ \mu T_t \ = \ B_{T_t}\,. \tag{52}$$

It can be shown [28] that $X_t$ is also a Lévy process, with a diffusion component $c\sigma$ and Lévy measure $\nu(x)$ given by

$$\nu(x) \ = \ \int_0^\infty \mathrm{d}t\, \wp(x, t)\rho(t)\,, \tag{53}$$

where $\wp(x, t)$ is the modelled by the normal probability density $\mathcal{N}(\mu t, t\sigma^2)$. This result is easy to understand; an intensity of $x$-sized jumps of the process $X$ is the probability that the process $B$ will in time $t$ change its value by $x$ multiplied by an intensity that the random time $T$ has the value $t$. Integration over all $t$'s then ensures that all random times compatible with the jump value $x$ are accounted for.

The formula (53) has formally the same structure as the defining formula for the canonical superstatistics — for each fixed $t$ the distribution of $X_t = B_{T_t}$ is a mixture of Gaussian distributions. One can even bolster this analogy by working with $\wp(x, t)$ and $\rho(t)$ that have substantially different characteristic scales. Unfortunately, the measures $\nu(x), \rho(t)$ are generally not probability measures. This can be rectified when Lévy processes in question have *finite activity* (i.e., finite average number of Poisson-type jumps per unite time), e.g. compound Poisson processes. In these cases Lévy measures $\nu = \nu(\mathrm{d}x)$ can be written as $\nu(\mathrm{d}x) = \lambda f(x)\mathrm{d}x$, where $\lambda$ is a finite constant representing an activity of jumps and $f(x)$ is a PDF determining the size of jumps [28]. A simple division by $\lambda$ then allows to pass to a superstatistics picture. This is not possible for Lévy processes with *infinite activity* (process has an infinite number of very small jumps in any finite time interval). In this case, it is no longer possible to separate the activity from the density $f(x)$ as both are non-trivially mixed in the Lévy measure $\nu(\mathrm{d}x)$. In such cases the superstatistics picture cannot be employed. Since one can pass from finite to infinite activity Lévy processes by a suitable change of parameters in the process, a potential breakdown of the canonical superstatistics can happen rather simply in the framework of Brownian subordination.

In passing we can mention that one can also construct Lévy processes from other Lévy processes (e.g., Lévy stable processes) by means of subordination (see, e.g., Ref. [53]). This offers a constructive way allowing to arrive at a non-canonical superstatistics where the prior $\wp(x,t)$ is some stable distribution.

## 6. Conclusions and outlooks

In this paper we have analyzed the issue of transmutation of statistics in the superstatistics framework. To illustrate the points involved we have used the high-frequency data for share-price log-returns of seven selected companies. We have first run a diagnostic check both with DFA and generalized Hurst exponents to identify the existence of well separated time-scales with distinct underlying dynamics. As a next step we did apply maximum likelihood method together with Kolmogorov–Smirnov, Cramér–von Mises and Anderson–Darling probability distances to fit optimally the two scale statistical behaviors with the three superstatistics universality classes. Within the seven analyzed companies we could positively identify four with clear signatures of the transition between two different superstatistics regimes. In particular, in all four aforementioned cases the short time scale (between 40 and 100 minutes) corresponded to the log-normal superstatistics while at large scales (around one trading day) the Gamma superstatistics appeared. The remaining three companies did not change superstatistics but remained locked within the log-normal superstatistics during the examined period. Due to numerical instabilities appearing in both kernel-density evaluations and in MFDFA at time scales shorter than 20 minutes, we have not been able to analyze minute-time scales and make comparison with findings of Ref. [18] on smallest time scales.

Another important issue, which we wanted to stress here is that the canonical superstatistics can *formally* appear (and as a rule it does) in a number of statistical contexts and one should be very careful when making unwarranted statements about affiliated dynamics. This point was illustrated with two examples, namely space-time (double) fractional diffusion and Brownian subordination. In both these situations the relevant PDF emerges in a form that mimics superstatistics marginal PDF (including the well separated scales), and yet for certain parameter space the would-be mixing distribution is negative or otherwise ill defined and the superstatistics picture is not applicable. This forewarning, however, does not intend to belittle a palpable importance of the superstatistics paradigm in hierarchical complex systems. It just aims at pointing some conceptual traps that any inexperienced practitioner should avoid.

Let us finally mention one interesting connection. As demonstrated, transmutation of statistics provides a convenient, easy-to-implement diagnostic tool for hierarchical stochastic processes with different (well separated) time-scales, which otherwise could not be easily distinguished and/or quantified. This ability to recognize within time series structures occurring at different time scales is similar in spirit to wavelet techniques. There one breaks time series into several sub-series which may be associated with a particular time scale which is then analyzed with ordinary time-series methods. One could thus think about a hybrid scheme where wavelet methods would represent a "lens" which could be used to zoom in on the details of a time series and superstatistics would then analyze each scale dynamics and draw an overall picture of the time series. Here, of course, an implicit assumption would need to be made that nested scale dynamics can be described via superstatistics universality classes. Work along these lines is presently under investigation.

## Acknowledgments

It is pleasure to acknowledge helpful conversations with D. Xu and H. Kleinert. P.J., J.K., M.P. and V.S. were supported by the Czech Science Foundation (GAČR) Grant No. 17-33812L. C.B. was supported by EPSRC via the grant EP/N013492/1.

## Appendix A: Glossary of companies used in text

In this appendix we provide a brief glossary of the companies whose share-price returns are considered in the main text.

| Company | | Sector | Stock exchange |
|---|---|---|---|
| AA | Alcola Inc. | basic materials | NYSE |
| KO | The Coca-Cola Company | consumer goods | NYSE |
| BAC | Bank of America Corporation | financial | NYSE |
| JNJ | Johnson & Johnson | healthcare | NYSE |
| GE | General Electric Company | industrial goods | NYSE |
| WMT | Wal-Mart Stories Inc. | services | NYSE |
| INTC | Intel Corporation | technology | NASDAQ |

Table 3. List of companies used in analysis

# Supplementary material: Transitions between superstatistical regimes: validity, breakdown and applications

Petr Jizba[a,b], Jan Korbel[c,a], Hynek Lavička[d,e], Martin Prokš[a], Václav Svoboda[a], Christian Beck[f]

[a]*Faculty of Nuclear Sciences and Physical Engineering, Czech Technical University in Prague, Břehová 7, 11519, Prague, Czech Republic*
[b]*Institute of Theoretical Physics, Freie Universität in Berlin, Arnimallee 14, 14195 Berlin, Germany*
[c]*Department of Physics, Zhejiang University, Hangzhou 310027, P. R. China*
[d]*Alten Belgium N.V., Chausse de Charleroi 112, 1060 Brussels, Belgium*
[e]*Department of Institutional, Environmental and Experimental Economics, University of Economics in Prague, Square W. Churchilla 4, 130 67 Praha 3, Czech Republic*
[f]*Queen Mary University of London, School of Mathematical Sciences, Mile End Road, London E1 4NS, United Kingdom*

**Abstract**

This supplementary material provides finer technical details to the analysis in the main text. We focus on dynamics of financial series for very short scales and finer technical details of multifractal detrended analysis. Finally, we present the results of MFDFA and superstatistical analysis for all investigated time series. We discuss the possible reasons leading to log-normal and gamma superstatistics and how it reflects in real financial series.

### Dynamics of log-returns for short scales below 20 minutes

Here we report the occurrence of an artificial structural form of kernel density which is likely due to very fine recording of data when liquidity of individual stocks, even of large companies used in analysis, is in question. The artificial structure is shown in Fig. 1 for company KO. The figure shows a kernel density estimate of probability density for the first 2000 log-returns at various time scales. Note that 2000 data points are clearly enough in order that the discrete levels could be considered significant. This artificial structure does occur also at different time positions in the time series, i.e. not necessarily at the beginning, and all studied time series possess the same flaw.

It is worth noting that discovery of this structure happened rather incidentally since the probability density estimate of log-returns for the whole series does not show any discreteness in log-returns. The explanation may be that the discrete levels are slightly shifted after some time period. From a physical point of view, we may say that for the minute scale the system has not yet reached equilibrium in the cell, i.e. assumption $\tau > \tau_r$, where $\tau_r$ is relaxation time of a cell, does not hold. One feasible solutions is to aggregate data into a higher time scale in which single stocks becomes liquid and log-returns continuous up to recording precision. By successive aggregation we see that a convenient sampling interval may be chosen to be 20 minutes since at this scale the artificial structure vanishes. It is worth of stressing that instabilities at the very same time scales were also observed in Section 3.1 in the main text, in connection with MFDFA analysis.

*Email addresses:* p.jizba@fjfi.cvut.cz (Petr Jizba), korbeja2@fjfi.cvut.cz (Jan Korbel), lavicka@fjfi.cvut.cz (Hynek Lavička), proks@fjfi.cvut.cz (Martin Prokš), svoboda@fjfi.cvut.cz (Václav Svoboda), c.beck@qmul.ac.uk (Christian Beck)

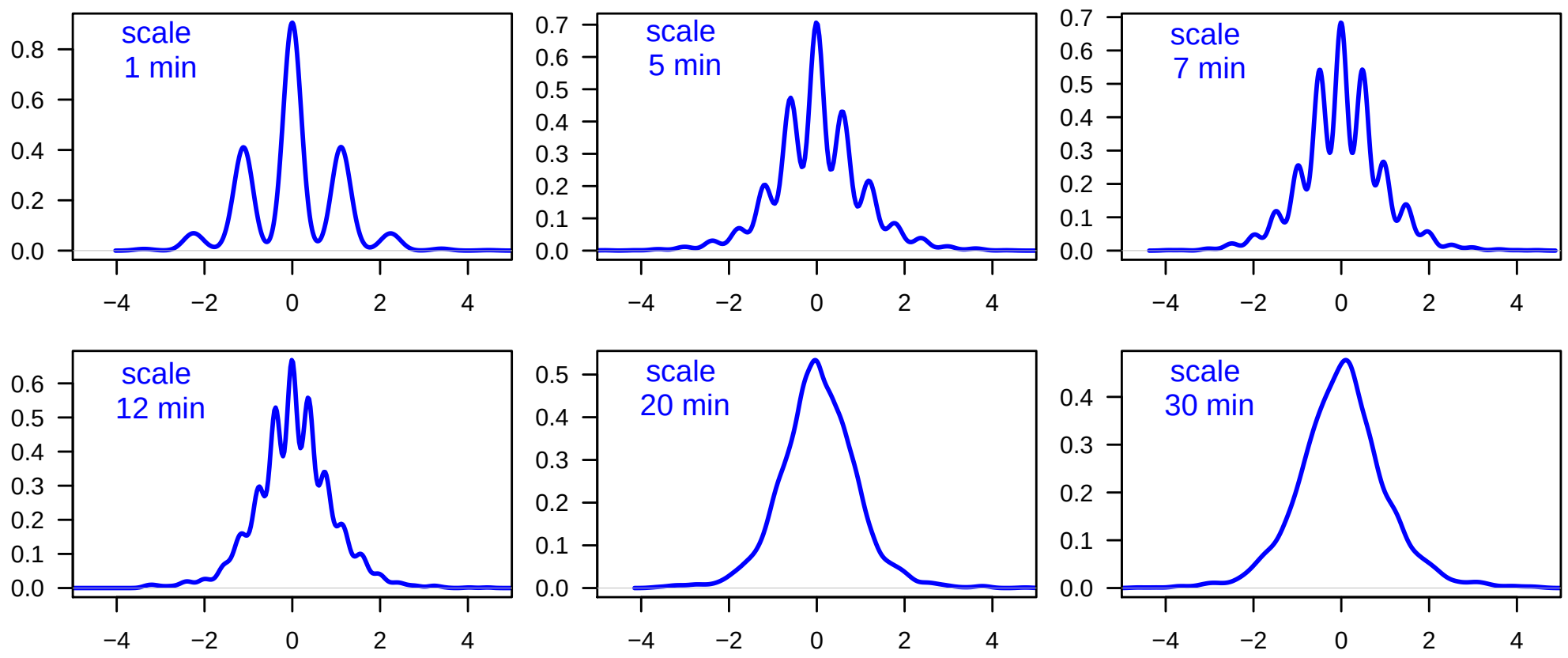

Figure 1. Artificial structure in log-returns time series. Example comes from company KO.

## Technical details of Multifractal detrended fluctuation analysis

This section discusses some technical details of the multifractal detrended fluctuation analysis. We discuss several steps that are necessary in order to prepare the data for the analysis. The first step is the pre-processing of the time-series. For the analysis, it is necessary that the time series is *stationary*. Unfortunately, not every investigated time series is stationary (financial market prices provide an example). In order to get the stationary series, we use the discrete derivation $D$ or integration $I$ of the series. This is realized as follows:

$$DX(t) = X(t) - X(t-1)$$

$$IX(t) = \sum_{i=0}^{t} X(i)$$

where $X(t)$ is the original series. This causes that for each operation the spectrum shifts to $\pm 1$. This is also advantageous if the spectrum is too shifted, as discussed in the main text. In this analysis, we cleaned up the raw dataset for mismeasurements and prepared the dataset as an input for MFDFA implementation. Results of the MFDFA analysis of the dataset exhibit single or two characteristic scales in the region of $[8, 2500]$ depending on the dataset. The first scale is caused by quantization of prices in combination with low liquidity. It is not present for all datasets. The process is governed by Poisson process rather diffusion. The latter regime is rather governed by diffusion. There is lower limit for a scale (depending on the type of asset) that approximates the process as a diffusive stochastic process. We note that preliminary results show existence of another scaling limit above 2500. We plan to investigate it in our next work. The second point which is worth to discuss is the order $o$ of the fitting polynomial. Typically, the most authors choose a very low order of the polynomial ($o = 1, 2$), because of computational purposes. On the other hand, for very long scales, the regions can contain more than one trend and then it is advantageous to use higher-order polynomials, which can deal with more trends.

## Transition between superstatistics regimes: financial time series

This section provides a more detailed analysis of the transition between superstatistical regimes for the financial time series. The dataset consists minute data of seven financial time series traded at New York SE and Nasdaq SE from various business sectors. The list of all investigated stocks is listed in Appendix A in the main text. Technical details of the analysis and comparison between particular distances has been extensively discussed in the main text, so we focus on the specific analysis for each series and the possible reasons leading to the superstatistics transition. In all cases, we observe that the short-scale dynamics is better described by log-normal superstatistics. It naturally describes multiplicative processes, such as geometric Brownian motion with fluctuating volatility or cascade processes.

On the other hand, Gamma superstatistics (also known as $\chi^2$ or Tsallis superstatistics) describing additive processes in fluctuating environment, outperforms the log-normal superstatistics for longer scales in some cases, but the transition region and the dynamics for short and long scales is very individual. We support the analysis by the MFDFA analysis for each series. We plot both the detrended fluctuation function and Anderson-Darling distance for each series for both short and long time scales and comment the behavior in Figs. 1-7. The table below summarizes optimal superstatistics for long and short time scales:

| **Company** | | **Short scale** | **Long scale** |
|---|---|---|---|
| AA | Alcola Inc. | log-normal | gamma |
| KO | The Coca-Cola Company | log-normal | gamma |
| BAC | Bank of America Corporation | log-normal | log-normal |
| JNJ | Johnson & Johnson | log-normal | log-normal |
| GE | General Electric Company | log-normal | log-normal |
| WMT | Wal-Mart Stories Inc. | log-normal | gamma |
| INTC | Intel Corporation | log-normal | gamma |

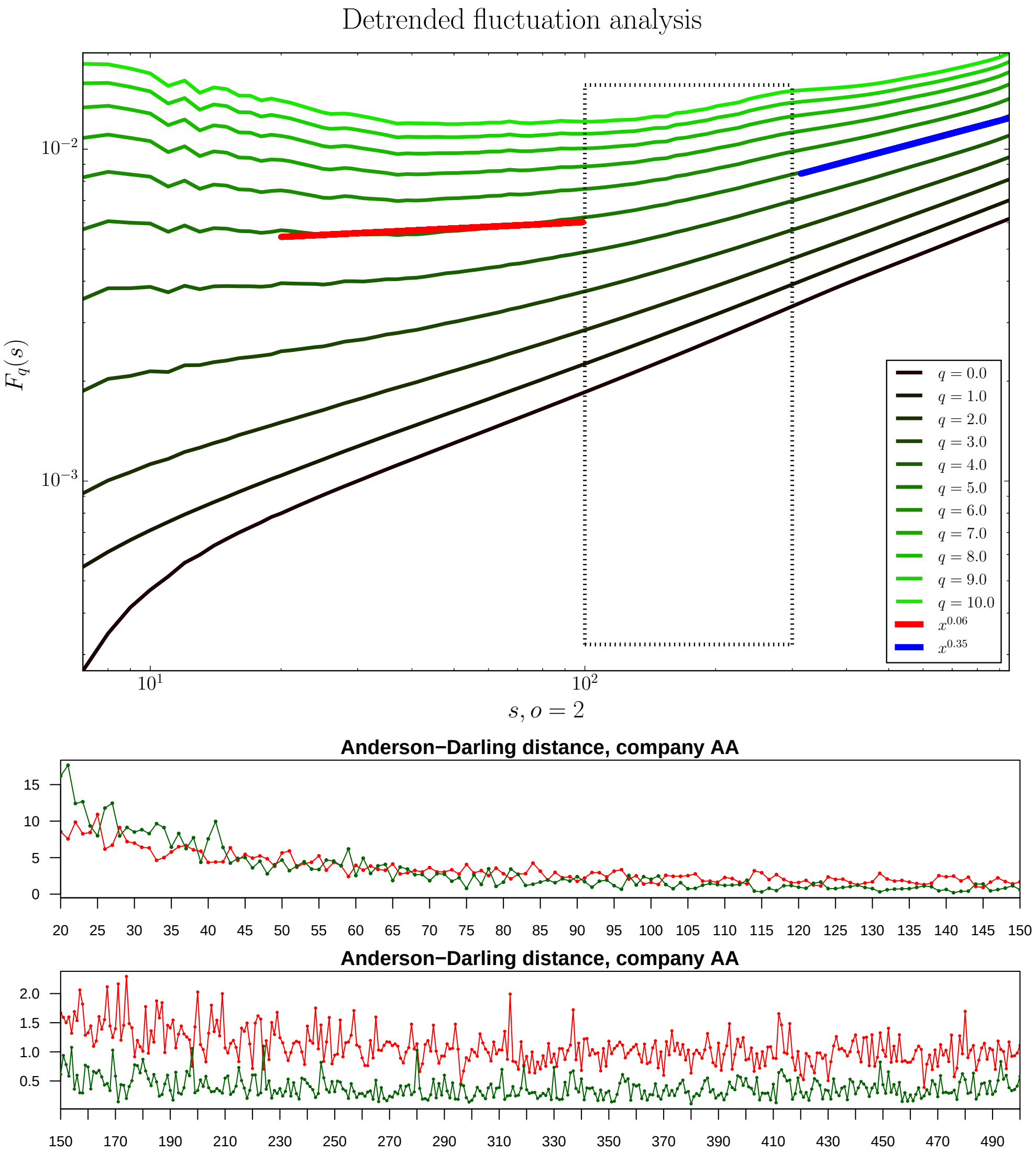

Figure 2. Statistical distance of superstatistical models for Alcoa time series. We observe a transition from log-normal superstatistics (red) on short scales to gamma superstatistics (green) on long scales. Transition scale is around 40-80 mins, but the transition regime is broad with no sharp border. This is also supported by DFA analysis, with two local scalings.

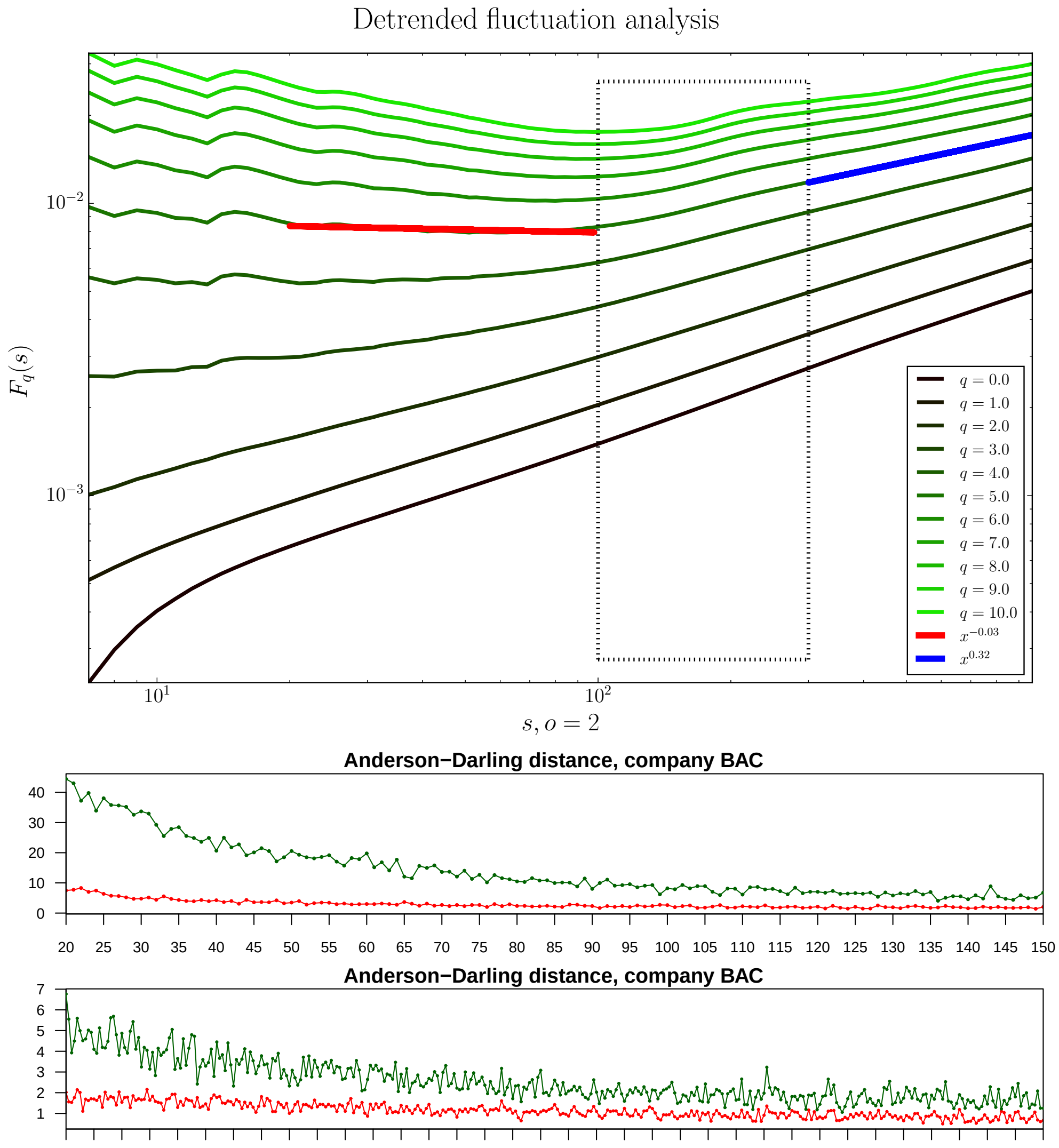

Figure 3. Statistical distance of superstatistical models for Bank of America time series. The gamma superstatistics (green) is remarkably outperformed by lognormal superstatistics (red) and we observe no transition, although we observe two local scalings in DFA analysis.

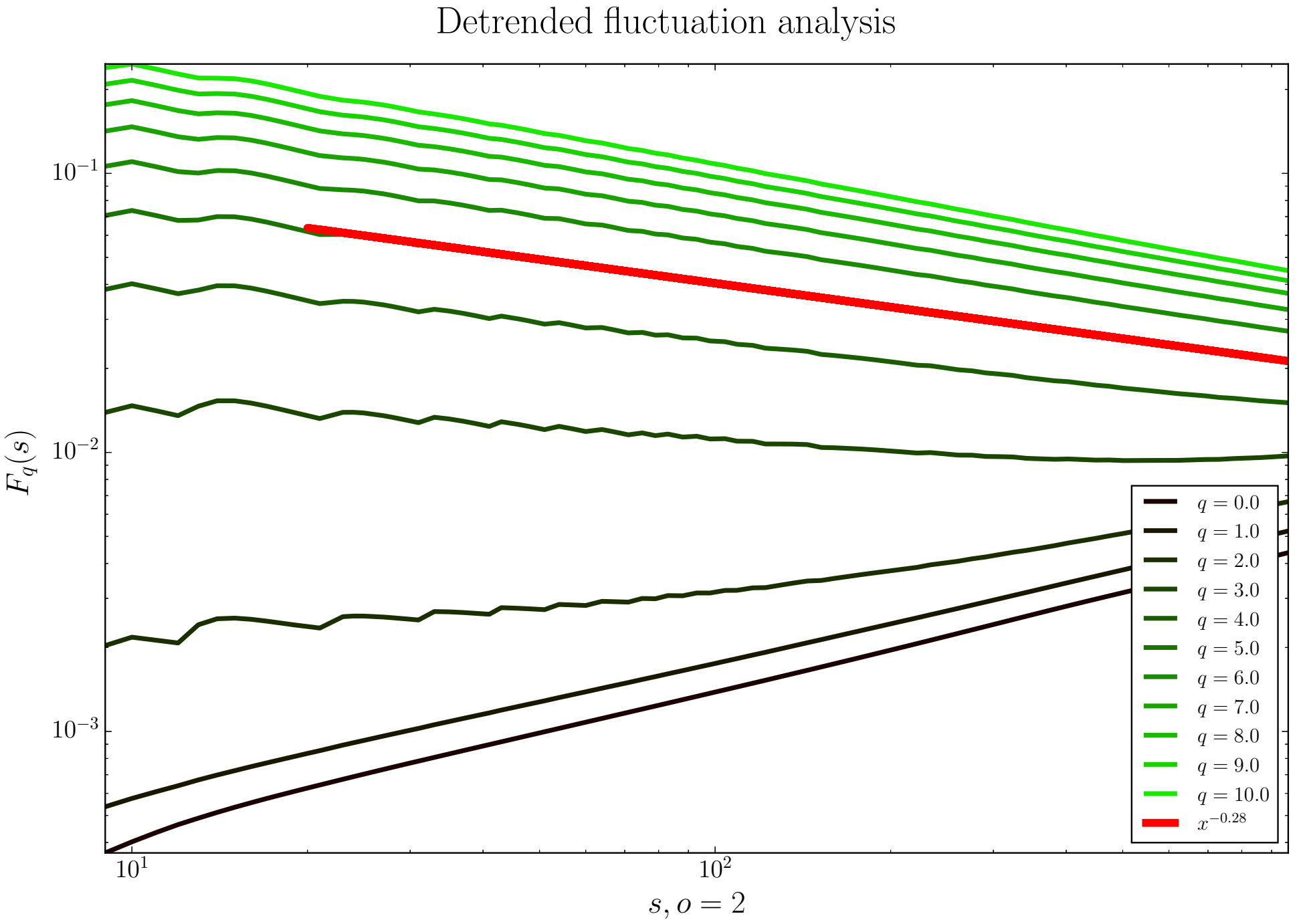

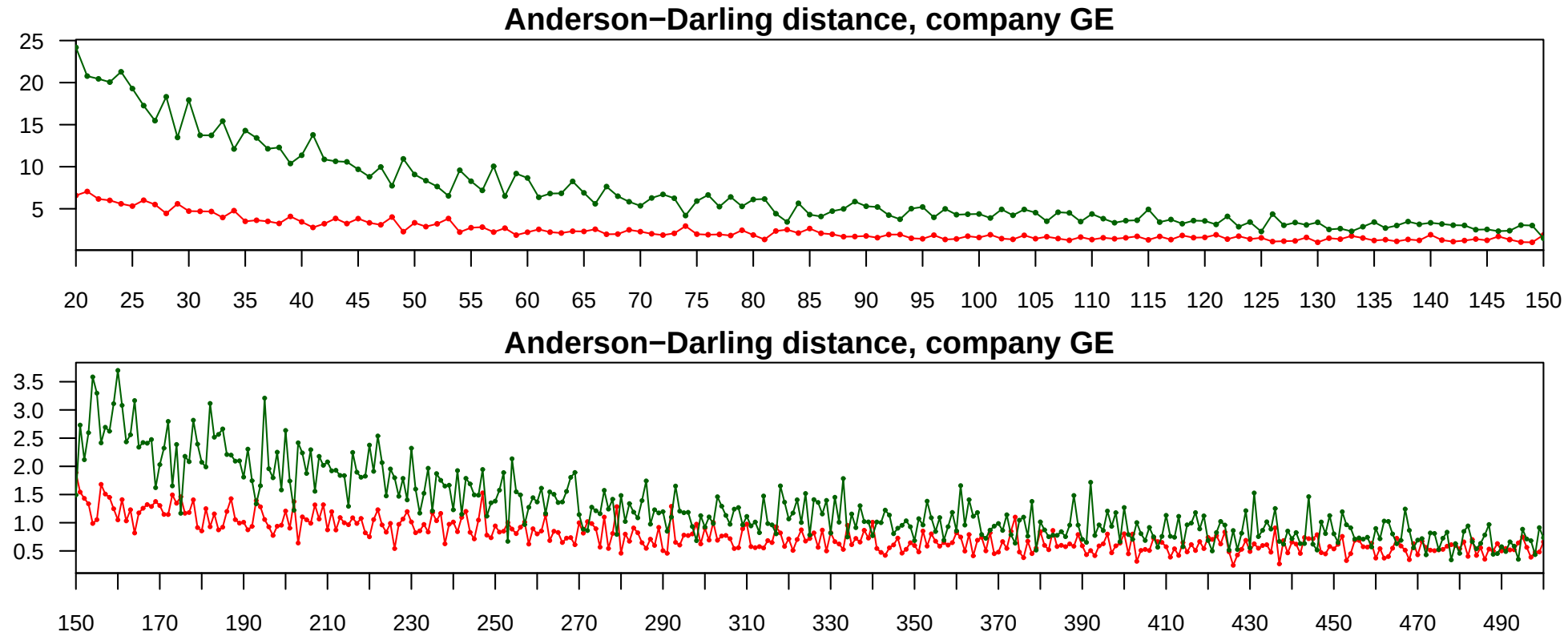

Figure 4. Statistical distance of superstatistical models for General Electric time series. The log-normal superstatistics (red) outperforms gamma superstatistics (green), but they become similar for long scales ( 500 mins).

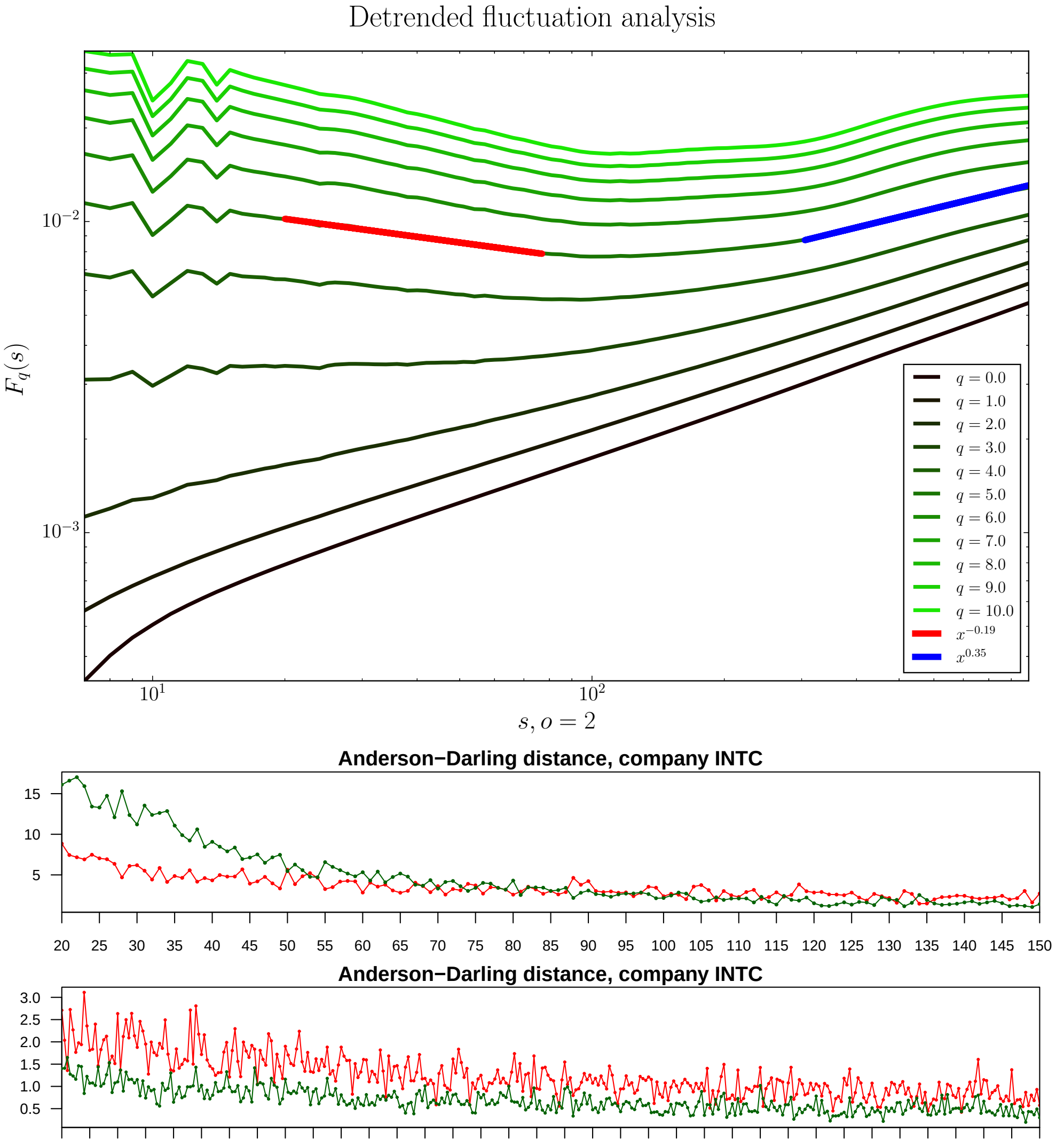

Figure 5. Statistical distance of superstatistical models for Intel Corp. time series. We can observe a transition from log-normal superstatistics (red) to gamma superstatistics (green). The transition regime is around 70-100 mins with no sharp border.

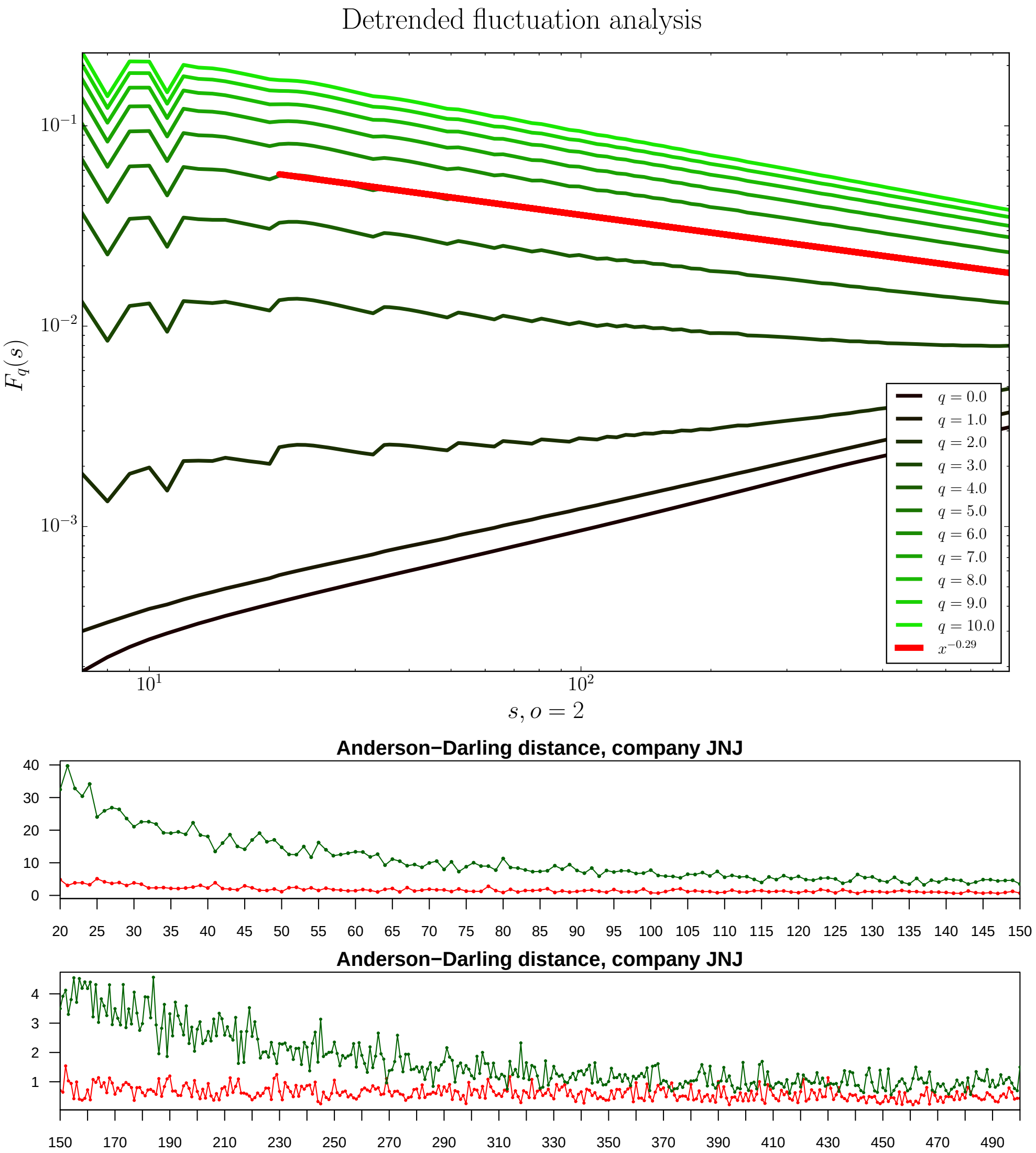

Figure 6. Statistical distance of superstatistical models for Johnson & Johnson time series. Again, log-normal superstatistics (red) is better than gamma superstatistics (green) on all investigated scales. This is also apparent from DFA analysis, where we observe single scaling for all investigated scales.

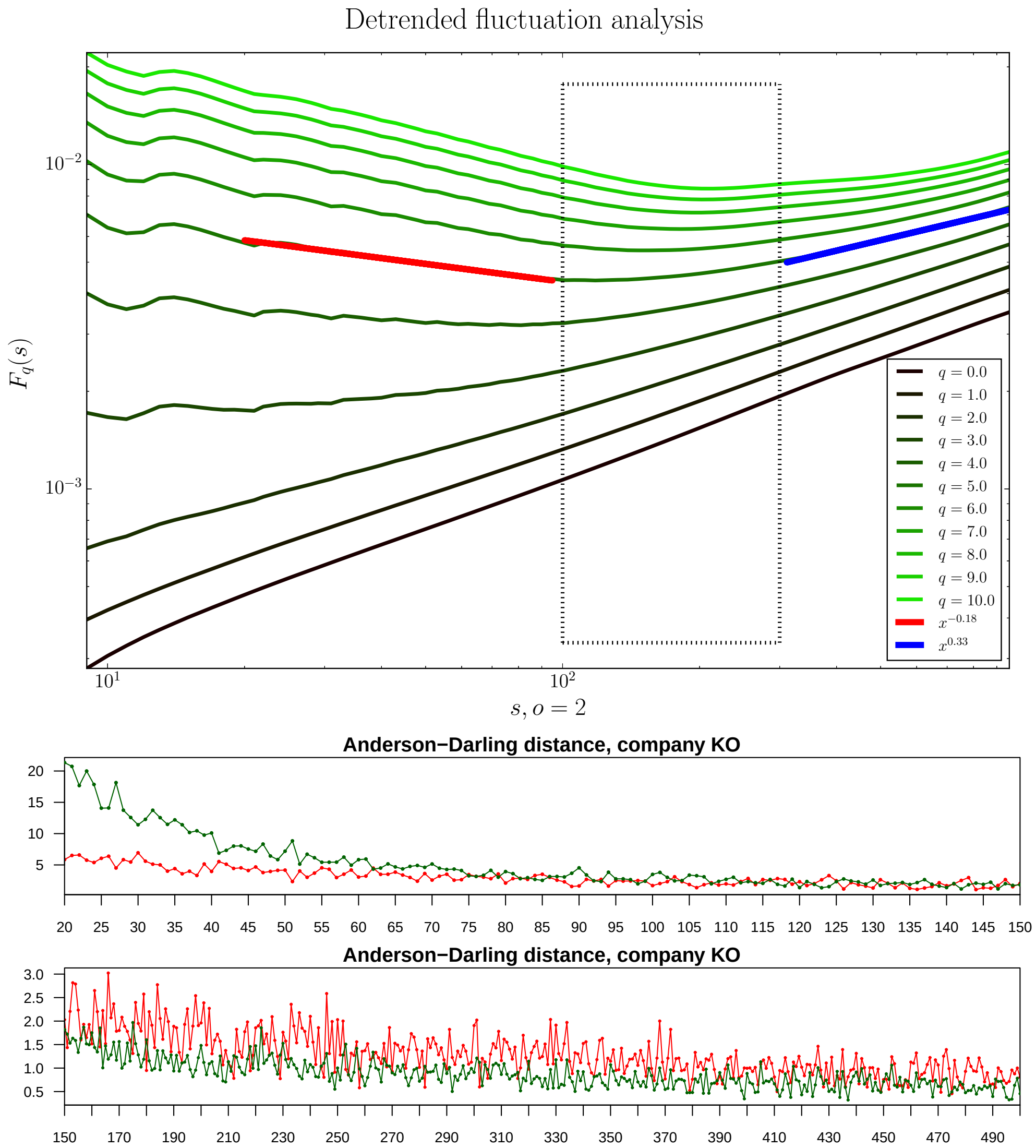

Figure 7. Statistical distance of superstatistical models for Coca-cola time series. We observe a transition between log-normal superstatistics (red) and gamma superstatstics (green). The transition regime is approximately around 80-120 mins. Compared to other time series, the transition is not so significant. This is also supported by DFA analysis, where we observe two distinct scaling rules with a smooth transition.

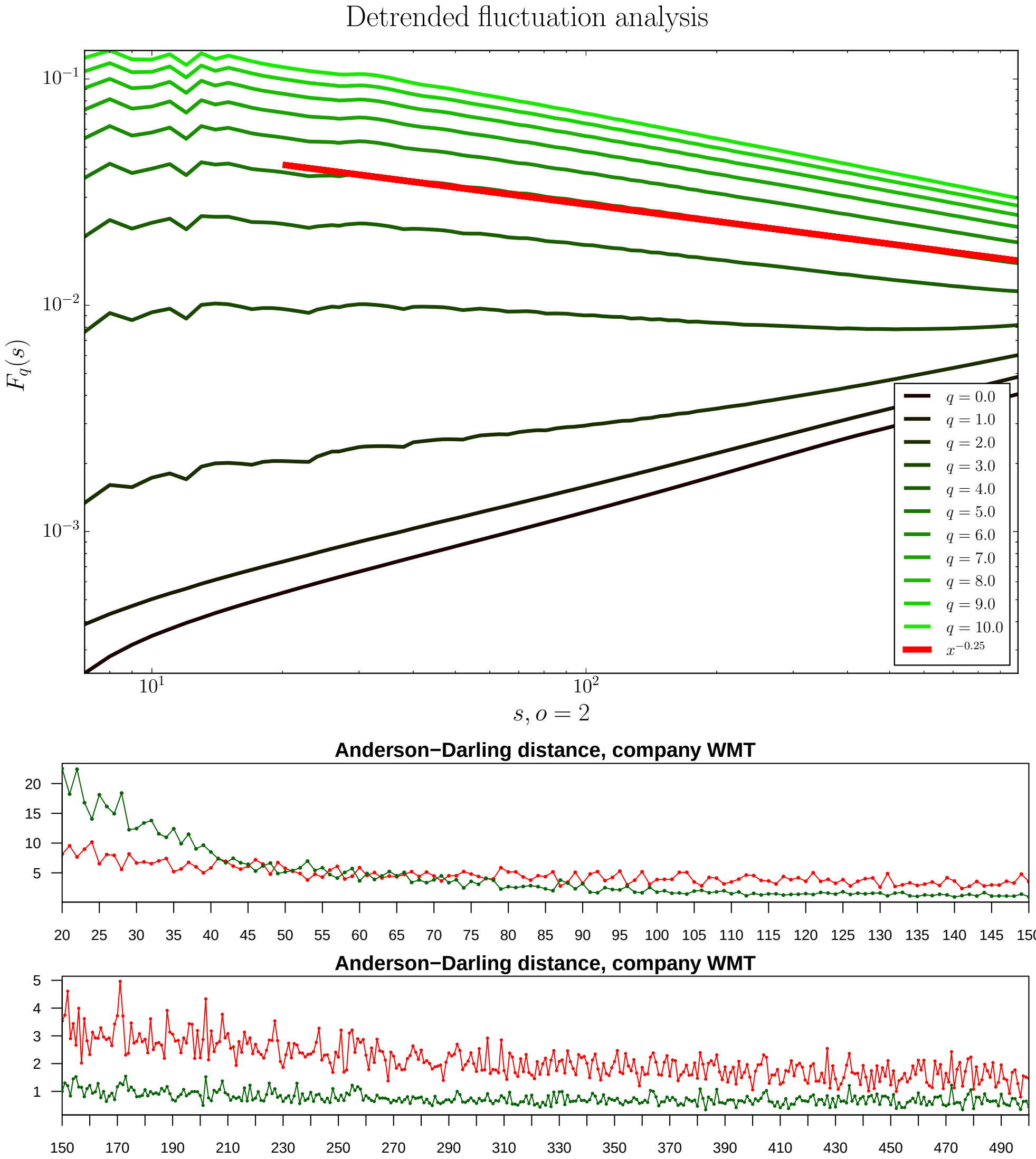

Figure 8. Statistical distance of superstatistical models for Wal-Mart time series. The transition between log-normal superstatistics (red) and gamma superstatistics (green) is remarkable, transition region is between 40-70 mins. In DFA analysis, we observe only one scaling rule.